\documentclass[final]{IEEEtran}
\usepackage{cite}
\usepackage{graphicx}
\usepackage{picinpar}
\usepackage[cmex10]{amsmath}
\usepackage{amsmath,amsfonts,amssymb}
\usepackage[justification=centering]{caption}

\usepackage{algorithmic}
\usepackage{subfigure}
\usepackage[ruled,lined,boxed]{algorithm2e}

\usepackage{stfloats}
\usepackage{bm}
\usepackage{color}
\usepackage{stfloats}
\usepackage{url}
\usepackage{caption} 
\usepackage{makecell}
\usepackage{multirow}
\usepackage{array}
\usepackage{booktabs}
\usepackage{geometry}
\usepackage{tabularx}
\newcolumntype{L}[1]{>{\raggedright\arraybackslash}p{#1}}

\begin{document}
	\pdfoptionpdfminorversion=6
	\newtheorem{lemma}{Lemma}
	\newtheorem{corol}{Corollary}
	\newtheorem{theorem}{Theorem}
	\newtheorem{proposition}{Proposition}
	\newtheorem{definition}{Definition}
	\newcommand{\e}{\begin{equation}}
		\newcommand{\ee}{\end{equation}}
	\newcommand{\eqn}{\begin{eqnarray}}
		\newcommand{\eeqn}{\end{eqnarray}}
	\renewcommand{\algorithmicrequire}{ \textbf{Input:}} 
	\renewcommand{\algorithmicensure}{ \textbf{Output:}} 
	
\title{E2E Learning Massive MIMO for Multimodal Semantic Non-Orthogonal Transmission and Fusion}

\author{Minghui Wu and Zhen Gao

\thanks{
	The work was supported in part by the Natural Science Foundation of China (NSFC) under Grant 62471036 and Grant U2233216, in part by Shandong Province Natural Science Foundation under Grant
	ZR2025QA30 and Grant ZR2022YQ62, in part by Beijing Natural Science Foundation under Grant L242011. \emph{(Corresponding author: Zhen Gao.)}	}
\thanks{Minghui Wu is with the School of Information and Electronics, Beijing Institute
	of Technology (BIT), Beijing 100081, China, and si also with State Key Laboratory of Environment Characteristics and Effects for Near-space, Beijing 100081, China. (e-mail: wuminghui@bit.edu.cn).} %
\thanks{Zhen Gao is with the State Key Laboratory of CNS/ATM, Beijing 100081, China, also with the MIIT Key Laboratory of Complex-Field Intelligent Sensing, Beijing 100081, China, also with BIT, Zhuhai 519088, China, also with the Advanced Technology Research Institute, BIT, Jinan 250307, China, and also with the Yangtze Delta Region Academy, BIT, Jiaxing 314019, China (e-mail: gaozhen16@bit.edu.cn).}

}
\maketitle
\thispagestyle{empty}
	
\begin{abstract}
This paper investigates multimodal semantic non-orthogonal transmission and fusion in hybrid analog-digital massive multiple-input multiple-output (MIMO). A Transformer-based cross-modal source-channel semantic-aware network (CSC-SA-Net) framework is conceived, where channel state information (CSI) reference signal (RS), feedback, analog-beamforming/combining, and baseband semantic processing are data-driven end-to-end (E2E) optimized at the base station (BS) and user equipments (UEs). 
CSC-SA-Net comprises five sub-networks: BS-side CSI-RS network (BS-CSIRS-Net), UE-side channel semantic-aware network (UE-CSANet), BS-CSANet, UE-side multimodal semantic fusion network (UE-MSFNet), and BS-MSFNet.
Specifically, we firstly E2E train BS-CSIRS-Net, UE-CSANet, and BS-CSANet to jointly design CSI-RS, feedback, analog-beamforming/combining with maximum {\emph{physical-layer's}} spectral-efficiency.
Meanwhile, we E2E train UE-MSFNet and BS-MSFNet for optimizing {\emph{application-layer's}} source semantic downstream tasks.  On these pre-trained models, we further integrate application-layer semantic processing with physical-layer tasks to  E2E train five subnetworks. 
Extensive simulations show that the proposed CSC-SA-Net outperforms traditional separated designs, revealing the advantage of cross-modal channel-source semantic fusion.


\end{abstract}
	

\IEEEpeerreviewmaketitle
	
\section{Introduction}\label{S1}

As we transition from 5G to the anticipated 6G networks, the demand for data transmission continues to escalate, driven by the proliferation of applications such as augmented reality, virtual reality, autonomous vehicles, and the Metaverse \cite{6G_intro1}. These applications generate vast amounts of data, challenging traditional communication systems. In this context, semantic communication emerges as a powerful solution, addressing the need for more efficient data transmission \cite{Sem_intro1, Sem_intro2, Sem_intro3}. By focusing on the transmission of meaningful information rather than raw data \cite{Sem_intro4}, semantic communication enables significant data compression, reducing the amount of data transmitted over the network. This approach is particularly valuable in communication scenarios where the goal is not merely to reconstruct data but to convey its underlying meaning \cite{JSAC_Wu3}, making it indispensable for future applications in 6G networks. 

Massive multiple-input multiple-output (MIMO) is a critical technology for future wireless communication systems \cite{MIMO1,MIMO2,MIMO3,MIMO4,MIMO5,new3,new6}. By leveraging large antenna arrays and advanced beamforming techniques, massive MIMO can significantly enhance transmission capacity, effectively addressing the growing data transmission demands in 6G networks. The ability to serve a large number of users simultaneously, while maintaining high throughput is one of the key advantages of massive MIMO. Integrating massive MIMO beamforming with semantic communication techniques is particularly promising. While massive MIMO increases system capacity through efficient beamforming, semantic communication helps reduce the transmission overhead by focusing on the most relevant information. This combination optimizes both the capacity of the system and the efficiency of data transmission, making it an ideal solution for the high data rate requirements of 6G communications.

\subsection{State-of the-Art}\label{S1.1}

In the field of semantic communication, the success of deep learning (DL) has sparked the adoption of deep neural network (DNN)-based architectures. In these systems, the transmitter, which performs semantic encoding, and the receiver, which performs semantic decoding, are treated as a pair of DNN-based autoencoders. The encoder at the transmitter semantically extracts the source data and encodes it into complex-valued transmission symbols, while the decoder at the receiver decodes the received symbols to reconstruct the data or perform other downstream tasks. These architectures have been widely applied in semantic communication systems to achieve better performance \cite{Sem_intro3}. Based on the type of task performed by the receiver, existing semantic communication works can be broadly classified into two categories: data reconstruction and task execution.

In the realm of data reconstruction, deep joint source-channel coding (DJSCC) for image data and its improved versions have been proposed in \cite{Deep-JSCC, NTSCC1, NTSCC2, ADJSCC}. In these methods, images are mapped by the semantic encoder to complex-valued transmission symbols, which are then reconstructed by the semantic decoder to enhance reconstruction performance. The DJSCC algorithm has been further improved for multi-user non-orthogonal multiple access (NOMA) transmission scenarios in \cite{DJSCC_NOMA}. Additionally, the DJSCC approach has been extended to semantic video transmission in \cite{DJSCC_video1, DJSCC_video2}. Furthermore, reference \cite{CDDM} utilized diffusion models to improve the quality of the received signals, thereby enhancing the image reconstruction performance in DJSCC.

For task execution applications, it is only necessary to extract and encode the semantic information relevant to the task. Specifically, \cite{R2_8} proposed a model for image retrieval tasks under power and bandwidth constraints, while \cite{R2_10} introduced a robust semantic communication system based on vector quantized variational autoencoders for image classification. Further, the authors of \cite{R2_11, R2_13} proposd multi-task semantic communication schemes. In \cite{sem-NOMA}, a multimodal semantic communication scheme was introduced that directly fuses the semantics of multiple UEs through non-orthogonal channel superposition for semantic segmentation tasks, thus reducing transmission overhead. Compared to data reconstruction applications, task execution applications can further reduce transmission overhead.

However, the aforementioned semantic communication schemes primarily focus on the extraction of semantics itself, without considering the impact of physical layer MIMO transmission. To this end, some studies have combined physical layer communication with semantic communication to better adapt to practical communication scenarios. Specifically, \cite{DJSCC_OFDM} explored the integration of orthogonal frequency division multiplexing (OFDM) with DJSCC, optimizing the end-to-end (E2E) performance of the entire semantic communication system. 
{References \cite{Vit_MIMO, CSI_fusion1, CSI_fusion2} further investigated the fusion of CSI with semantics in MIMO, enhancing the transmission of semantic features. {These works are confined to $2{\times}2$ small-scale MIMO under {perfect CSI} at both transmitter and receiver, and they rely on conventional precoding/equalization,  in which case this fusion can effectively handle semantic transmission weights between different data streams.} While these studies considered the integration of MIMO and semantic communication, they only use MIMO’s CSI to assist semantic encoding, while still employing traditional MIMO physical-layer designs. {\cite{MIMO_sem_E2E} further improves semantic transmission via decoupled pre-training and deep unfolding, yet it also assumes {perfect CSI} and thus adopts a WMMSE-style precoder.} Additionally, \cite{JSAC_Wu3} further integrated massive MIMO precoding design with multi-user semantic communication, enhancing the quality of massive MIMO image transmission, {but it does not embed the CSI acquisition  (e.g., CSI-RS design and feedback) into an E2E semantic-communication architecture.}}


In massive MIMO–OFDM systems, the channel estimation, channel feedback, and precoding design critically determine physical‐layer performance. Classical compressed sensing (CS)‑ or codebook‑based channel estimation and feedback schemes, as well as iterative beamforming algorithms, often incur substantial CSI reference signals (CSI-RS) and feedback overhead and on‑line computational burden \cite{JSAC_Wu1}. To address these challenges, recent works have increasingly leveraged DL techniques across these three modules to enhance accuracy while reducing latency and overhead under constrained resources. 

For channel estimation, the authors in \cite{Ma_tvt} proposed a convolutional neural network (CNN)‑based channel estimator for hybrid massive MIMO systems, while \cite{CE1_JSAC} introduced a model‑driven DL solution. In channel feedback, csiNet and its variants \cite{csinet1,csinet2,csinet3} employed CNNs to compress and reconstruct downlink CSI features, effectively reducing feedback bits without sacrificing reconstruction fidelity. In precoding design, the authors of \cite{R1-12, R1-13, R2-13} proposed the use of supervised CNNs to approximate traditional iterative beamforming, which has been further extended to hybrid beamforming in massive MIMO systems in \cite{major-1}. Transfer learning strategies further adapted beamforming networks to diverse channel conditions \cite{R2-12}. Reference \cite{WY_Trans} validated the efficacy of Transformer models for channel estimation, feedback, and precoding in massive MIMO–OFDM systems. Furthermore, recent studies have unified CSI-RS design, CSI acquisition, and precoding into a single E2E neural network, optimizing it jointly for spectral efficiency  \cite{JSAC_Wu1,My1-46,My1-48,E2E-Pre2}. These approaches obviated the need for explicit channel estimation and channel feedback, and achieves substantial communication capacity under extremely low CSI-RS and feedback overhead.

\begin{table*}[t]
	
	\centering
	\captionsetup{font={color = {black}}}
	\caption{Comparison with representative semantic communication works.}
	\label{tab:semcomm_comparison}
	\resizebox{\textwidth}{!}{
		\begin{tabular}{lccccccc}
			\toprule
			\textbf{Reference} & \textbf{MIMO size} & \textbf{Link direction} & \textbf{\#UEs} & \textbf{CSI acquisition} & \textbf{Beamforming} & \textbf{Architecture} & \textbf{Downstream task} \\
			\midrule
			\cite{Vit_MIMO}          & $2\times 2$  & Downlink & Single-UE & No  & No  & Digital & Image reconstruction \\
			\cite{CSI_fusion1}       & $2\times 2$  & Downlink & Single-UE & No  & No  & Digital & Image reconstruction \\
			\cite{CSI_fusion2}       & $2\times 2$  & Uplink   & Multi-UEs & No  & No  & Digital & Image reconstruction \\
			\cite{MIMO_sem_E2E}      & $16\times 8$ & Uplink   & Multi-UEs & No  & Yes & Digital & Image classification \\
			\cite{JSAC_Wu3}          & $64\times 1$ & Downlink & Multi-UEs & No  & Yes & Digital & Image reconstruction \\
			\textbf{Our work}        & $64\times 8$ & Uplink   & Multi-UEs & Yes & Yes & \textbf{Hybrid} & \textbf{Semantic segmentation} \\
			\bottomrule
		\end{tabular}
	}
	\vspace{-0.5em}
\end{table*}

\subsection{ Motivation and Contribution}\label{S1.2}

Although \cite{Vit_MIMO,CSI_fusion1,CSI_fusion2} have investigated semantic communication schemes in MIMO systems, they rely on conventional physical‑layer techniques such as singular‑value decomposition (SVD)‑based beamforming, which require precise CSI. Even though \cite{MIMO_sem_E2E,JSAC_Wu3} proposed joint designs of the MIMO physical layer and semantic communication, these approaches assume CSI as perfectly known and neglect both the acquisition process and Doppler effects. In the context of massive MIMO–OFDM systems, constraints on limited CSI-RS and feedback overhead inevitably lead to inaccurate channel estimation, posing serious challenges to system performance.


The discussion in the previous subsection on DL-based MIMO physical-layer design demonstrates that E2E communication frameworks can significantly enhance the capacity of massive MIMO systems under stringent resource constraints. {Meanwhile, non-orthogonal multiple access has been extensively studied for massive connectivity and signal detection in massive MIMO systems~\cite{new1,new2,new4,new5}, where non-orthogonal transmission has been shown to yield performance gains over orthogonal schemes under limited time–frequency resources.} Furthermore, {building on this line of thought,} \cite{sem-NOMA} showed that in multi-user transmit, single-user receive semantic communication scenarios—where multiple users serve the same semantic task—the non-orthogonal superposition of users’ signals at the receiver can be interpreted as multi-user semantic fusion. This strategy avoids the high overhead of orthogonal transmission methods and reduces communication cost. Inspired by these insights, we combine the E2E design methodology with non-orthogonal semantic transmission to propose a unified framework for MIMO physical-layer design and semantic transmission in massive MIMO systems with multi-user transmission and single-user reception, i.e., cross-modal source-channel semantic-aware network (CSC-SA-Net).
{We list the main differences in scenarios between the proposed semantic transmission scheme and existing similar schemes in Table \ref{tab:semcomm_comparison}.}
The contributions of this work are summarized as follows:

\begin{enumerate}
	\item \textbf{BS-side CSI-RS network (BS-CSIRS-Net) and UE/BS-side channel semantic-aware network (UE/BS-CSANet).} We introduce the BS-CSIRS-Net and UE/BS-CSANet that jointly design analog precoding and combining matrices, embedding CSI-RS design, channel feedback, and beamforming in a single E2E architecture to enhance physical‑layer transmission efficiency. {Within this architecture, the learned channel semantics also support non-orthogonal superposition across cooperating UEs, enabling over-the-air aggregation that converts interference into task-useful statistics and aligns the physical layer with the downstream objective.}
	
	\item \textbf{UE-side multimodal semantic fusion network (UE-MSFNet).} We propose the UE-MSFNet, which fuses the channel semantic feature with the source semantic feature to map transmit symbols implicitly, eliminating explicit demodulation reference signals (DMRS) and avoiding throughput loss from non‑data reference signals. The adaptive mapping, based on the combined source semantic feature and channel semantic feature derived from the CSI-RS, functions as an implicit DMRS allocation mechanism, facilitating adaptation to changing CSI conditions.
	
	\item \textbf{BS-side multimodal semantic fusion network (BS-MSFNet).} We present the BS-MSFNet to merge multiple UEs’ source and channel semantic features into a unified representation at the BS, obviating per‑user explicit detection and focusing on task‑relevant fused semantics to maximize end‑to‑end semantic communication efficiency. { Under non-orthogonal superposition, BS-MSFNet performs over-the-air semantic fusion, which yields higher task-level utility than separately reconstructing each user’s stream; our OMA baseline confirms consistent gains at comparable budgets (especially under low transmit symbol budgets).}
	
	\item \textbf{Task‑driven three-stage training.} We integrate BS-CSIRS-Net, UE/BS-CSANet, and UE/BS-MSFNet into the cross-modal source-channel semantic-aware network (CSC-SA-Net) framework and perform three-stage training. In the first stage, UE/BS-MSFNet are pretrained on the downstream task for robust initialization. In the second stage, BS-CSIRS-Net and UE/BS-MSFNet are jointly optimized for spectral efficiency. Finally, the entire CSC-SA-Net is jointly trained on the downstream task. By aligning downlink CSI-RS design, CSI processing, beamforming, and semantic transmission with the downstream task objective, our approach achieves higher semantic accuracy under resource-limited conditions.''

\end{enumerate}

\begin{figure*}[!t]
	\vspace*{-2mm}
	\centering
	\includegraphics[width = 2\columnwidth,keepaspectratio]{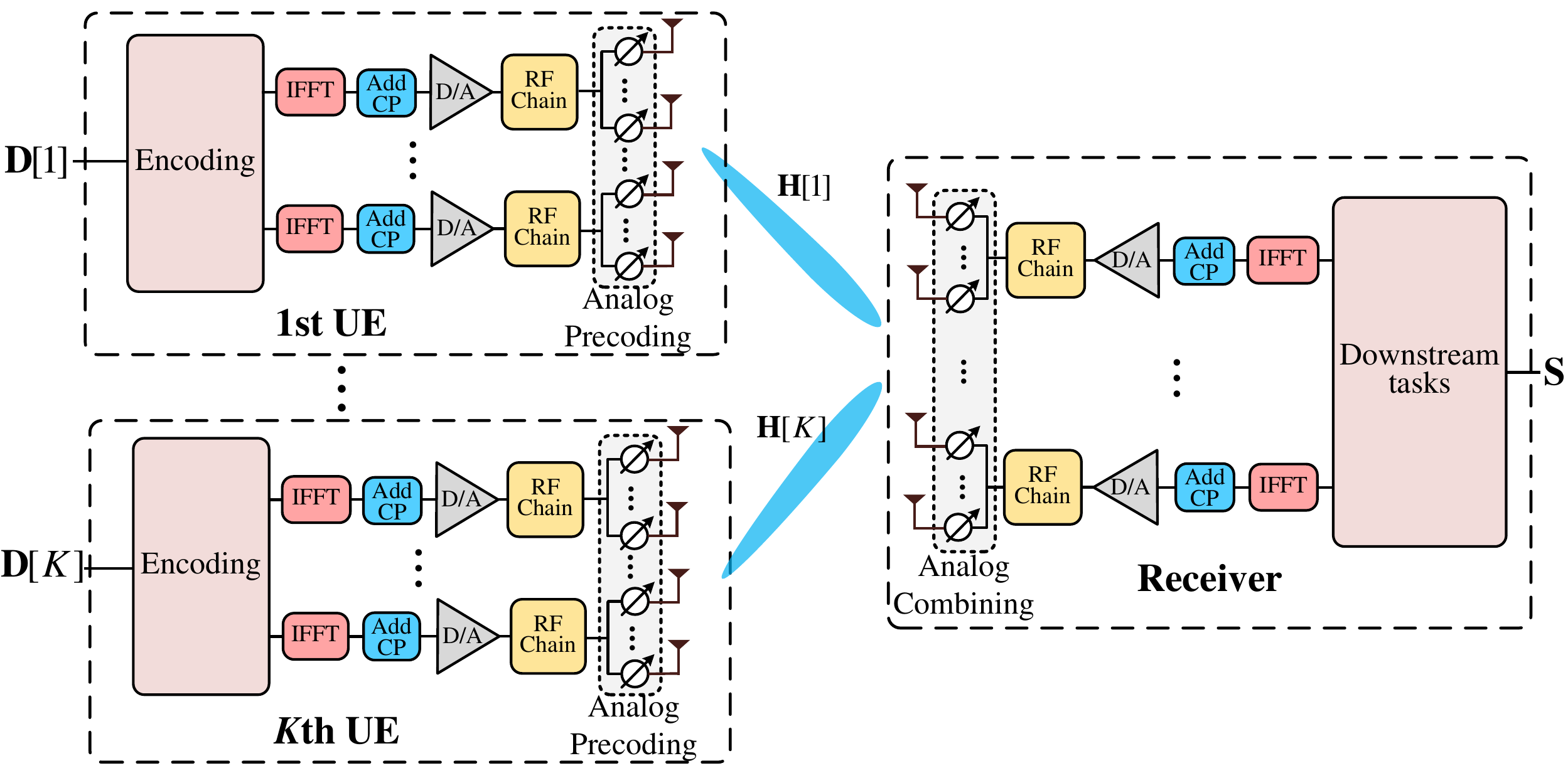}
	\captionsetup{font={footnotesize}, singlelinecheck = off, justification = raggedright,name={Fig.},labelsep=period}
	\caption{Multi-user massive MIMO uplink semantic transmission system diagram.}
	\label{fig:SYS} 
	\vspace*{-2mm}
\end{figure*}

\textit{Notation}: We use lower-case letters for scalars, lower-case boldface letters for column vectors, and upper-case boldface letters for matrices.
Superscripts $(\cdot)^*$, $(\cdot)^{\rm T}$, $(\cdot)^{\rm H}$, $(\cdot)^{-1}$ and $(\cdot)^\dagger$ denote the conjugate, transpose, conjugate transpose, inversion and Moore-Penrose inversion operators, respectively.
$\left\| {\mathbf{A}} \right\|_F$ is the Frobenius norm of ${\mathbf{A}}$.
${\rm{vec}}( {\mathbf{A}} )$ and ${\rm{angle}}( {\mathbf{A}} )$ denote the vectorization operation and the phase values of ${\mathbf{A}}$, respectively.
${\mathbf{I}_n}$  denotes the $n\times n$ identity matrix, while $\bm{1}_n$ ($\bm{0}_n$) denotes the vector of size $n$ with all the elements being $1$ ($0$).
$\Re\{\cdot\}$ and $\Im\{\cdot\}$ denote the real part and imaginary part of the corresponding argument, respectively.
$[\mathbf{A}]_{m,n}$ denotes the $m$th-row and $n$th-column element of $\mathbf{A}$, while $[\mathbf{A}]_{[:,m:n]}$
is the sub-matrix containing the $m$th to $n$th columns of $\mathbf{A}$. The expectation is denoted by $\mathbb{E}(\cdot)$. $\frac{{\partial a}}{{\partial b}}$ is the partial derivative of $a$ with respect to $b$.


\section{System Model}\label{sec:SYS}
We consider the uplink semantic transmission in a multi-user massive MIMO-OFDM system, as shown in Fig.~\ref{fig:SYS}. In this system, $K$ user equipments (UEs), each equipped with a hybrid analog-digital MIMO array with $N_t$ antennas and $N_{{\rm RF},t}$ RF chains, transmit semantic information over $N_c$ subcarriers to a base station (BS) equipped with a hybrid MIMO array with $N_r$ antennas and $N_{{\rm RF},r}$ RF chains.

The source data $\mathbf{D}[k]$ of the $k$th UE is first encoded into a complex-valued baseband signal ${{\bf{S}}_{{\rm{BB}}}}[k] = \big[ {{\left[ {{{\bf{s}}_{{\rm{BB}}}}[k,1,1], \cdots ,{{\bf{s}}_{{\rm{BB}}}}[k,1,{N_c}]} \right]}^T}; \cdots ;[ {{\bf{s}}_{{\rm{BB}}}}[k,Q,1], \cdots , $ ${{\bf{s}}_{{\rm{BB}}}}[k,Q,{N_c}] ]^T \big] \in\mathbb{C} {^{Q \times {N_c} \times {N_{{\rm{RF}},{{t}}}}}}$, where ${\bf{s}}_{{\rm{BB}}}[k,q,{n}]\in\mathbb{C} {^{{N_{{\rm{RF}},{{t}}} \times {1} }}}$ is the baseband signal on the $n$th subcarrier of the $q$th OFDM symbol. This process can be expressed as \footnote{Note that we have not defined the dimensions or provided detailed explanations for $\mathbf{D}[k]$ and $\mathbf{R}$ in \eqref{equ:enc} and \eqref{equ:dec}; these will be given in detail in the next section.}
\begin{equation} \label{equ:enc}
	\mathbf{S}_{\rm BB}[k] = \mathcal{F}_{\rm enc}\bigl( \mathbf{D}[k] \bigr),
\end{equation}
where $\mathcal{F}_{\rm enc}(\cdot)$ denotes the encoding function. Subsequently, analog precoding is applied to form the antenna-domain transmit signal
\begin{equation}
	\begin{split}
		\mathbf{s}[k,q,n]
		&= \mathbf{F}_{\mathrm{RF}}[k]\,\mathbf{s}_{\mathrm{BB}}[k,q,n]\in\mathbb{C} {^{{N_{{{t}}} \times {1} }}},\\
		&{\rm for} \quad n=1,\ldots,N,\;\;q=1,\ldots,Q,
	\end{split}
\end{equation}
where $\mathbf{F}_{\rm RF}[k]\in\mathbb{C}^{N_t\times N_{{\rm RF},t}}$ is the analog precoding matrix and $\mathbf{s}[k,q,n]$ is the transmit signal on the $n$th subcarrier of the $q$th OFDM symbol. Since the analog precoder is implemented using RF phase shifters, it must satisfy a constant modulus constraint, i.e., 
$\bigl|\left[\mathbf{F}_{\rm RF}[k]\right]_{i,j}\bigr|=1,\quad \forall\, i,j.$
In addition, the transmit signal is subject to a power constraint, i.e., 
$\|\mathbf{s}[k,q,n]\|_F^2\leq P_t,$
where $P_t$ denotes the transmit power\footnote{Note that we do not specify a separate digital precoding module because the encoding block in Fig.~\ref{fig:SYS} is defined as the entire process that maps the source signal to the transmit baseband  signal; the digital precoding operation can be regarded as part of this process. This encoding module may be implemented in a modular manner using conventional source-channel coding, modulation, and baseband digital precoding, or it may be designed in an E2E joint fashion.}. 

After transmission over the channel, the signals from the $K$ UEs are non-orthogonal superimposed at the BS receive antennas. Specifically, the received signal on the $n$th subcarrier of the $q$th OFDM symbol is given by
\begin{equation} \label{equ:rec1}
	\mathbf{y}[q,n]=\sum_{k=1}^K \mathbf{H}[k,q,n]\, \mathbf{s}[k,q,n] + \mathbf{n}[q,n]\in\mathbb{C}^{N_r\times 1},
\end{equation}
where $\mathbf{H}[k,q,n]\in\mathbb{C}^{N_r\times N_t}$ denotes the channel from the $k$th UE to the BS on the $n$th subcarrier during the $q$th OFDM symbol, and $\mathbf{n}[q,n]$ is the noise vector. Then, the BS applies an analog combiner to obtain the received baseband  signal
\begin{equation} \label{equ:rec2}
	\mathbf{y}_{\rm BB}[q,n] = \mathbf{W}_{\rm RF}\,\mathbf{y}[q,n]\in\mathbb{C}^{N_{{\rm{ RF}},r}\times 1},
\end{equation}
where $\mathbf{W}_{\rm RF}\in\mathbb{C}^{N_{{\rm RF},r}\times N_r}$ is the analog combining matrix, which also satisfies the constant modulus constraint 
\[
\bigl|\left[\mathbf{W}_{\rm RF}\right]_{i,j}\bigr|=1,\quad \forall\, i,j.
\]
By aggregating the baseband received  signals across all subcarriers and OFDM symbols, the overall baseband received signal $\mathbf{Y}_{\rm BB}\in\mathbb{C}^{N_c\times N_{{\rm RF},r}\times Q}$ is obtained. Finally, the BS performs semantic decoding on $\mathbf{Y}_{\rm BB}$ and executes subsequent tasks to produce the final output $\mathbf{R}$, i.e.,
\begin{equation} \label{equ:dec}
	\mathbf{R} = \mathcal{F}_{\rm dec}\bigl( \mathbf{Y}_{\rm BB} \bigr),
\end{equation}
where $\mathcal{F}_{\rm dec}(\cdot)$ denotes the decoding function.

\begin{figure*}[!t]
		\vspace*{-2mm}
	\centering
	\includegraphics[width = 2\columnwidth,keepaspectratio]{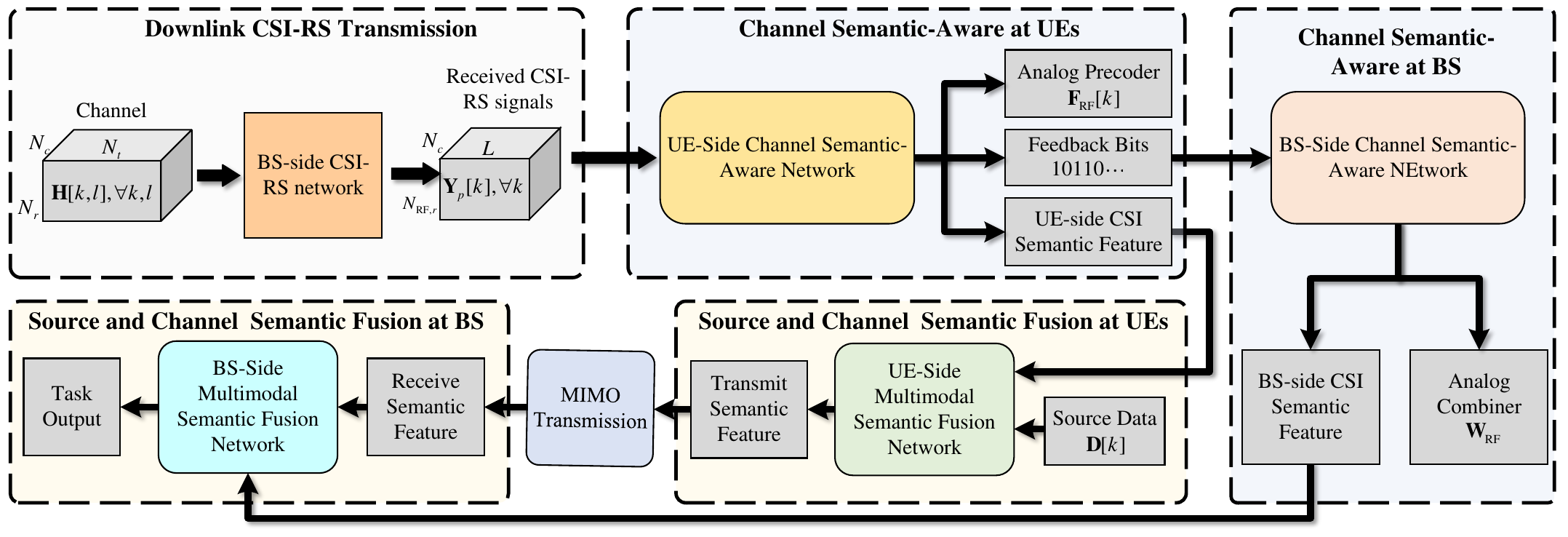}
	\captionsetup{font={footnotesize}, singlelinecheck = off, justification = raggedright,name={Fig.},labelsep=period}
	\caption{The proposed CSC-SA-Net for E2E multimodal semantic non-orthogonal transmission and fusion in massive MIMO, which composed of five modules: 1) the BS-CSIRS-Net for generating the BS-side CSI-RS to enable channel semantic extraction; 2) the UE-CSANet for extracting the UE-side channel semantic feature from the received CSI-RS, producing feedback bits, and performing analog precoding; 3) BS-CSANet for recovering the channel semantic feature at the BS and constructing the analog combiner; 4) the UE-MSFNet for extracting the source semantic feature, fusing them with the UE-side channel semantic feature, and generating the transmit semantic feature; 5) the BS-MSFNet for fusing the over-the-air non-orthogonal superposition of multimodal semantic features with the BS-side channel semantic feature to perform the final downstream task.}
	\label{fig:E2E} 
		\vspace*{-2mm}
\end{figure*}

We consider a typical time-varying multipath massive MIMO-OFDM channel. In particular, the channel is assumed to be constant within each OFDM symbol and changes from one OFDM symbol to the next. Let \(L_p[k]\) denote the number of multipath components between the $k$th UE and the BS. The channel matrix on the \(n\)th subcarrier for the \(q\)th OFDM symbol is modeled as{{\footnote{{In this work, we employ theoretical channel models in order to provide controllable and reproducible evaluation, and to ensure fair comparison with existing baselines. These models allow us to clearly illustrate the performance trends of the proposed framework under varying conditions. While practical channel datasets such as those generated by Sionna could provide further validation, incorporating them is left as an important direction for future work.}}}}
\begin{equation}\label{eq:channel_model}
	\begin{split}
		\mathbf{H}[k,q,n] = \frac{1}{\sqrt{L_p[k]}} \sum_{l=1}^{L_p[k]} \alpha[{l,k}]\, \mathbf{a}_r\big(\theta_r[{l,k}]\big) \mathbf{a}_t^H\big(\theta_t[{l,k}]\big)  \\
		\times e^{-\textsf{j}\frac{2\pi n\tau[{l,k}]}{N_c T_s}}\, e^{\textsf{j}2\pi f_{d}[l,k]\,q\,T_I},
	\end{split}
\end{equation}
where \(\alpha[{l,k}]\) is the complex gain of the \(l\)th path, \(\theta_t[{l,k}]\in[-\pi,\pi]\) and \(\theta_r[{l,k}]\in[-\pi,\pi]\) denote the angle of departure (AoD) at the UE and the angle of arrival (AoA) at the BS, respectively, \(\tau[{l,k}]\) is the delay of the \(l\)th path, \(f_{d}[l,k]\) is the Doppler frequency of the \(l\)th path, \(T_s\) is the OFDM sampling interval, and \(T_I\) is the time interval of one OFDM symbol.

For the UE equipped with a half-wavelength uniform linear array (ULA) of \(N_t\) elements, the transmit array steering vector is given by
\begin{equation}\label{eq:steering_tx}
	\mathbf{a}_t(\theta_t)=\frac{1}{\sqrt{N_t}}\left[1,\, e^{\textsf{j}\frac{2\pi}{\lambda}d\sin(\theta_t)},\, \ldots,\, e^{\textsf{j}\frac{2\pi}{\lambda}d\,(N_t-1)\sin(\theta_t)}\right]^T,
\end{equation}
where \(\lambda\) is the wavelength and \(d=\lambda/2\) is the spacing between adjacent antenna elements.

Similarly, for the BS equipped with a half-wavelength ULA of \(N_r\) elements, the receive array steering vector is expressed as
\begin{equation}\label{eq:steering_rx}
	\mathbf{a}_r(\theta_r)=\frac{1}{\sqrt{N_r}}\left[1,\, e^{\textsf{j}\frac{2\pi}{\lambda}d\sin(\theta_r)},\, \ldots,\, e^{\textsf{j}\frac{2\pi}{\lambda}d\,(N_r-1)\sin(\theta_r)}\right]^T.
\end{equation}

\section{Proposed Scheme}\label{sec:Pro}
In this section, we first present the processing procedure of the proposed E2E joint CSI acquisition, beamforming, and semantic transmission and fusion scheme, i.e., CSC-SA-Net, in an uplink time division duplex (TDD) multi-user hybrid analog-digital massive MIMO-OFDM system, as shown in Fig. \ref{fig:E2E}{\footnote{{Semantic in this paper refers to {learned, task-oriented latent features} rather than human-interpretable symbols or raw pixels/bits. Concretely, we use (i) \emph{source semantics}: high-level representations extracted from RGB–thermal inputs that encode class cues, boundaries, and contextual relations for the downstream task; and (ii) \emph{channel semantics}: compact features distilled from CSI-RS and feedback that summarize propagation/array conditions for adaptive analog precoding/combining and implicit symbol mapping. These latents are {jointly} optimized E2E for the downstream objective, and are transmitted over the air; we do {not} transmit raw CSI matrices or pixel/bit streams.}}}. In the following, we elaborate on how to model the physical layer design and semantic transmission problem as an E2E neural network.

\subsection{Processing Procedure and Problem Formulation}
For the specific semantic task, we consider the multi-user semantic segmentation task described in \cite{sem-NOMA, MFNet}. As outlined in \cite{sem-NOMA}, in a cloud-edge collaborative (or edge computing) architecture, infrared sensors and cameras on vehicles collect environmental data. These multimodal data are compressed by a dedicated neural network encoder and then transmitted to the cloud/edge server in the form of features through the wireless channel. The cloud/edge server utilizes its powerful computational resources to jointly process the multimodal features, performing semantic segmentation to assist in path planning and decision-making.

To fully leverage massive MIMO to enhance semantic transmission efficiency, we need to extract useful information from the channel that can aid in physical layer design, namely channel semantic information, and perform beamforming. 
In TDD mode with a hybrid analog-digital array at both the BS and the UEs, estimating uplink high-dimensional CSI from a limited number of RF chains is a non-trivial problem and requires multiple downlink CSI-RS OFDM symbols and uplink feedback bits. In the downlink CSI-RS training stage, we assume that the BS transmits $L$ orthogonal CSI-RS OFDM symbols. The CSI-RS signal received by the $k$-th UE on the $n$-th subcarrier and $l$-th OFDM symbol can be expressed as
\begin{equation}\label{equ:CSI-RS}
	\begin{split}
		{{\bf{Y}}_p}[k,l,n] &= {{\bf{V}}_{{\rm{RF}}}}[l]{\bf{H}}[k,l]{{\bf{X}}_{{\rm{RF}}}}[l]{{\bf{x}}_{{\rm{BB}}}}[l,n] \\
		&\quad + {\bf{z}}[k,l,n] \in {^{{N_{{\rm{RF}},{{t}}}} \times 1}},
	\end{split}
\end{equation}
where ${\bf{z}}[k,l,n] \in \mathbb{C} {^{{N_{{\rm{RF}},{{t}}}} \times 1}}$ represents AWGN, ${{\bf{X}}_{{\rm{RF}}}}[l]\in \mathbb{C} {^{N_r \times {N_{{\rm{RF}},{{r}}}} }}$ and ${{\bf{x}}_{{\rm{BB}}}}[l,n]\in \mathbb{C} {^{{N_{{\rm{RF}},{{r}}}} \times 1}}$ are the analog RF and digital baseband CSI-RS signals transmitted by the BS, and ${{\bf{V}}_{{\rm{RF}}}}[k,l]$ is the UE's analog RF CSI-RS combining matrix. Note that since we employ a fully connected hybrid MIMO architecture, the CSI-RS matrix is divided into digital and analog components, where the analog CSI-RS must satisfy the constant modulus constraint, i.e., $\bigl|\left[\mathbf{X}_{\rm RF}[l]\right]_{i,j}\bigr|=1,\quad \forall\, i,j.$ and $\bigl|\left[\mathbf{V}_{\rm RF}[k,l]\right]_{i,j}\bigr|=1,\quad \forall\, i,j.$ The transmitted CSI-RS signal must also satisfy the transmit power constraint, i.e., $\|{{\bf{X}}_{{\rm{RF}}}}[l]{{\bf{x}}_{{\rm{BB}}}}[q,n]\|_F^2\leq P_t$. By combining the received signals across all subcarriers and CSI-RS OFDM symbols, the received CSI-RS signal for the $k$-th user can be written as ${{\bf{Y}}_p}[k] \in {^{{N_c} \times {N_{{\rm{RF}},{{t}}}} \times L}}$.

Next, we consider the $k$th UE's extraction of channel semantic information from ${{\bf{Y}}_p}[k] \in {^{{N_c} \times {N_{{\rm{RF}},{{t}}}} \times L}}$ to design precoding, while also compressing the channel semantic information into $B$ bits and feeding it back to the BS. This process can be mathematically expressed as
\begin{equation}
	\left\{ {{\bf{q}}[k],{{\bf{F}}_{{\rm{RF}}}}[k]} \right\} = {\cal P}({{\bf{Y}}_p}[k]),
\end{equation} 
where ${\mathbf q}[{k}] \in {^{B \times 1}}$ is the $k$-th UE's binary bit vector fed back to the BS, and ${\mathcal P}(\cdot)$ is a mapping function that maps the received CSI-RS signal $\widetilde{\mathbf Y}_p[{k}]$ to the feedback bit vector ${\mathbf q}[{k}]$ and beamformer ${{\bf{F}}_{{\rm{RF}}}}[k]$.

Based on the feedback bits from all $K$ users, the BS designs the analog combiner, which can be expressed as
\begin{equation}
	{{\bf{W}}_{{\rm{RF}}}} = \mathcal{C}({\bf{q}}[1], \cdots ,{\bf{q}}[K]),
\end{equation} 
where ${\mathcal C}(\cdot)$ is a mapping function from the received feedback bit vectors to the analog combiner ${{\bf{W}}_{{\rm{RF}}}}$.

Subsequently, the $K$ UEs begin the uplink semantic transmission as described in Section \ref{sec:SYS}. In the multi-user semantic segmentation task considered in this section, the number of UEs is 2. The first UE’s source is a three-channel RGB image $\mathbf{D}[1]\in\mathbb{C}^{3\times H\times W}$, and the second UE’s source is a single-channel infrared image $\mathbf{D}[2]\in\mathbb{C}^{1\times H\times W}$, where $H=480$ and $W=640$ are the height and width of the image. The output of the semantic segmentation is $\mathbf{R}\in\mathbb{C}^{9\times H\times W}$, which represents the probability of the semantic segmentation class for each pixel. Our ultimate goal is to maximize the accuracy of the semantic segmentation. To address this, we model the above process as an E2E neural network, as shown in Fig. \ref{fig:E2E}.

\subsection{Downlink CSI-RS Transmission} 
At the downlink CSI-RS transmission stage, as represented in (\ref{equ:CSI-RS}), the parameters to be designed include the digital baseband CSI-RS signals ${{\bf{x}}_{{\rm{BB}}}}[l,n]$, analog RF CSI-RS signals ${{\bf{X}}_{{\rm{RF}}}}[l]$, and the analog CSI-RS combiner ${{\bf{V}}_{{\rm{RF}}}}[l]$ for $1 \leq n \leq N_c$ and $1 \leq l \leq L$. Similar to the learnable CSI-RS design proposed in \cite{JSAC_Wu2},
we model the entire downlink CSI‐RS transmission process as a learnable network, denoted {BS-CSIRS-Net}, whose input is the channel realization $\mathbf H[k,l],\forall k,l$ and whose output is the received CSI-RS $\mathbf Y_p[k,l,n],\forall k,l,n$. We treat the digital baseband CSI-RS signals ${{\bf{x}}_{{\rm{BB}}}}[l,n]$, the phase values $ \mathbf{P}_{{\rm RF}}[l] \in \mathbb{R} {^{N_r \times {N_{{\rm{RF}},{{r}}}} }}$  of the analog RF CSI-RS signals, and the phase values $ \mathbf{Q}_{{\rm RF}}[l] \in \mathbb{R} {^{ {N_{{\rm{RF}},{{t}}}} \times N_t}}$ of the analog CSI-RS combiner as trainable parameters. These matrices can be learned and optimized during the DL training stage. We then apply the complex exponentiation function directly to the phase matrices, ensuring that the unit modulus constraints for both the analog CSI-RS signals and the analog CSI-RS combiner are satisfied, which can be expressed as

\begin{equation}
	\mathbf{X}_{{\rm RF}}[l] = \exp\left( {\rm j}  \mathbf{P}_{{\rm RF}}[l] \right),
\end{equation}
and
\begin{equation}
	\mathbf{V}_{{\rm RF}}[l] = \exp\left( {\rm j}  \mathbf{Q}_{{\rm RF}}[l] \right).
\end{equation}

Finally, we address the power normalization of the digital CSI-RS signals to comply with the power constraint, given as
\begin{equation}
	{{\bf{x}}_{{\rm{BB}}}}[l,n] = \frac{{\sqrt {{P_t}} {{\bf{x}}_{{\rm{BB}}}}[l,n]}}{{{{\left\| {{{\bf{X}}_{{\rm{RF}}}}[l]{{\bf{x}}_{{\rm{BB}}}}[l,n]} \right\|}_F}}},\forall l,n.
\end{equation}

\begin{figure*}[!t]
	\vspace*{-2mm}
	\centering
	\includegraphics[width = 2\columnwidth,keepaspectratio]{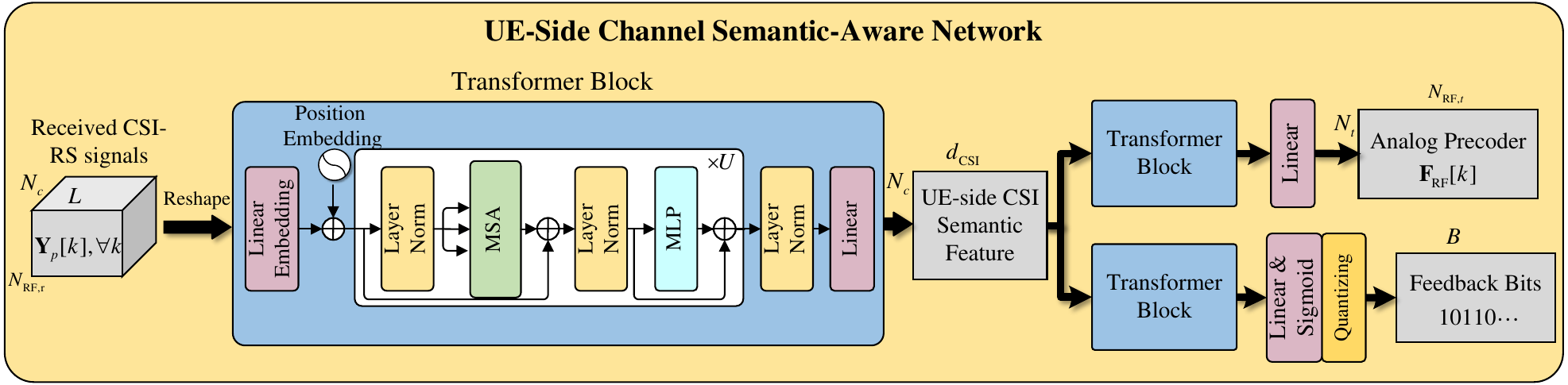}
	\captionsetup{font={footnotesize}, singlelinecheck = off, justification = raggedright,name={Fig.},labelsep=period}
	\caption{Network Architecture of the Proposed UE-CSANet.}
	\label{fig:CSI_Sem_UE} 
	\vspace*{-2mm}
\end{figure*}

\subsection{Channel Semantic-Aware at UEs}
After the UEs receive the CSI-RS signals, they need to extract useful channel semantic information and design the analog precoder accordingly. At the same time, the semantic information is compressed into a bit vector, which is then fed back to the BS for the design of the analog combiner. This process is modeled as the Transformer-based UE-CSANet, as shown in Fig. \ref{fig:CSI_Sem_UE}.

It is worth noting that the Transformer structure in DL has been extensively applied in fields such as natural language processing and computer vision \cite{Transformer,VIT}. It has been shown to outperform fully connected neural networks, CNNs, and other architectures in various scenarios. Unlike the convolutional operations in CNNs \cite{csinet1}, which can only extract features from local areas, the self-attention mechanism in Transformers is capable of capturing global features. This allows the Transformer to globally extract the inter-subcarrier correlation of the input signal, providing corresponding weighting coefficients for each subcarrier's components, which enhances performance.

The standard Transformer accepts a 1D real-valued sequence as input and produces a 1D real-valued sequence as output \cite{Transformer,VIT}. In order to process the input of the proposed compressor, which is a 3D complex-valued matrix ${\mathbf Y}_p[k]$ with dimensions $N_c \times N_{{\rm RF},t} \times L$, we first reshape the received CSI-RS signal into a 2D complex-valued matrix $\tilde{\mathbf Y}_{p}[k] \in \mathbb{R}^{N_c \times N_{{\rm RF},t}L}$ by combining the last two dimensions. Then, we concatenate the real and imaginary parts to generate a 1D real-valued sequence  with dimension $N_c \times  2N_{{\rm RF},t}L$
\begin{equation}
	\begin{cases}
		\ \ \ \ 	\left[ \bar{\mathbf Y}_{p}[k]\right]_{[:,1:N_{{\rm RF},t}L]}=\Re\{{\tilde{\mathbf Y}_{p}[k]}\},\\
		\left[ \bar{\mathbf Y}_{p}[k]\right]_{[:,N_{{\rm RF},t}L+1:2N_{{\rm RF},t}L]}=\Im\{{\tilde{\mathbf Y}_{p}[k]}\},
	\end{cases}
\end{equation}
where the number $N_c$ of subcarriers serves as the effective input sequence length for the Transformer.

In the Transformer, the input sequence is first transformed into a sequence of vectors with dimension $d_{\rm model}$ by using a fully connected linear embedding layer, followed by a position embedding layer. The position embedding uses a sine function of varying frequencies to represent the positions of different subcarriers. Then, the Transformer applies $U$ identical layers to extract the features of the input sequence. Each layer consists of a multi-head self-attention sublayer and a multilayer perceptron (MLP) sublayer. Finally, the extracted features are processed by a fully connected linear layer to output the UE-side channel semantic feature $\mathbf{S}_{\rm CSI, UE}[k]$ with dimension $N_c \times d_{\rm CSI}$.

The subsequent processing is divided into two branches. In the upper branch, the channel semantic feature is input into a Transformer, where the subcarrier dimension $N_c$ of the channel semantic feature is treated as the sequence length for the Transformer input. The output from this Transformer is then passed through a fully connected linear layer to generate the phase values ${\mathbf \Theta}_{\rm RF}[k] \in \mathbb{R}^{N_t \times N_{{\rm RF},t}}$ of the analog precoding matrix. By applying the complex exponential function to the phase matrix ${\mathbf \Theta}_{\rm RF}[k]$, the {UE-CSANet} produces an analog precoding matrix that satisfies the unit modulus constraint, i.e.,
\begin{equation}
	{{\bf{F}}_{{\rm{RF}}}}[k] = \exp \left({\rm{j}}{{\bf{\Theta }}_{{\rm{RF}}}}[k]\right), \quad 1 \le k \le K.
\end{equation}

In the lower branch, the channel semantic feature is similarly input into another Transformer, after which the sequence is compressed into a codeword through a fully connected linear layer followed by a sigmoid activation function. Finally, the compressed codeword is quantized into $B$ feedback bits by a quantization layer.{{\footnote{For the quantization and dequantization layer, we first normalize the input features to a range of zero to one using the sigmoid function. Then, we divide this interval into four equally spaced intervals, and each feature scalar is quantized to the center of its interval and represented with two bits. At the receiving end, we dequantize the received bits to obtain the quantization value corresponding to each feature. Note that in the process of deep learning training, quantization and dequantization do not have gradients, so conventional backpropagation cannot be directly performed. For this purpose, we adopted the direct gradient propagation method, which uses the straight-through gradient estimation method to directly copy and transfer the gradient of the dequantized features to the features before quantization during the gradient backpropagation process.}} }

This entire process can be summarized as
\begin{equation}
	\left\{ {{\bf{q}}[k],{{\bf{F}}_{{\rm{RF}}}}[k],{{\bf{S}}_{{\rm{CSI}},{\rm{UE}}}}[k]} \right\} = \mathcal{P}({{\bf{Y}}_p}[k];{W_{{\rm{UE - CSANet}}}}),
\end{equation}
where ${\mathcal W}_{\rm UE-CSANet}$ denotes the learnable parameters of the neural network, and ${\cal P}(\cdot;{\mathcal W}_{\rm UE-CSANet})$ is the mapping function that transforms the received CSI-RS signals ${\mathbf Y}_{\rm p}[k]$ into the feedback vector ${\mathbf q}[k]$, the analog precoding matrix ${{\bf{F}}_{{\rm{RF}}}}[k]$, and the UE-side channel semantic feature $\mathbf{S}_{\rm CSI, UE}[k]$. Note that we take the UE-side channel semantic feature $\mathbf{S}_{\rm CSI, UE}[k]$ as part of the network output here, rather than just an intermediate variable. This is because in the subsequent source semantic transmission, we can further utilize it for fusion to alleviate the impact of non-ideal channel characteristics on transmission.

\begin{figure*}[!t]
	\vspace*{-2mm}
	\centering
	\includegraphics[width = 2\columnwidth,keepaspectratio]{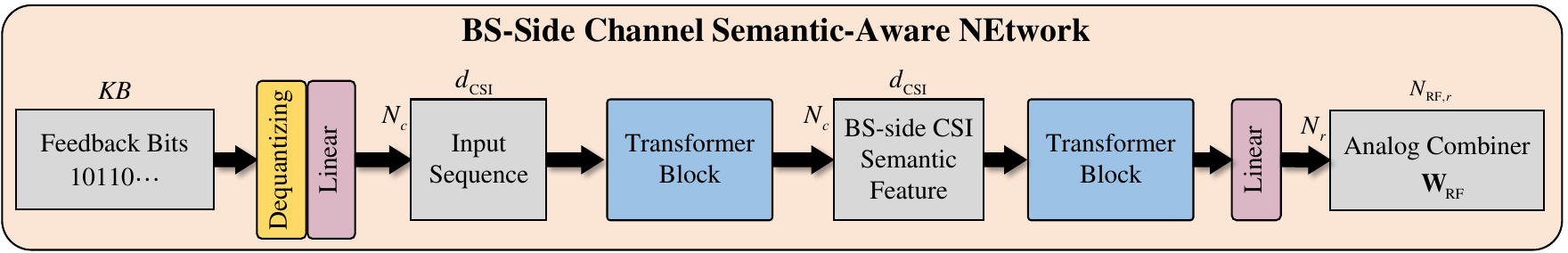}
	\captionsetup{font={footnotesize}, singlelinecheck = off, justification = raggedright,name={Fig.},labelsep=period}
	\caption{Network Architecture of the Proposed BS-CSANet.}
	\label{fig:CSI_Sem_BS} 
\end{figure*}

\subsection{Channel Semantic-Aware at BS}

The BS needs to extract useful channel semantic information from the feedback bit vectors received from all $K$ UEs and design the analog combiner accordingly. We assume that the feedback from the UEs to the BS is error-free. This process is modeled as the BS-CSANet, as shown in Fig. \ref{fig:CSI_Sem_BS}. The feedback bit vectors from all $K$ UEs are first concatenated into a vector of size $KB$, and this concatenated vector is then passed through a dequantization layer. After dequantization, the resulting sequence is processed by a fully-connected linear layer to convert it into the input sequence for the Transformer. The Transformer then extracts features from the input sequence and outputs the BS-side channel semantic feature $\mathbf{S}_{\rm CSI, BS}$ with dimensions $N_c \times d_{\rm CSI}$. 

Next, the channel semantic feature is input into another Transformer, and the phase values ${\mathbf \Phi}_{\rm RF} \in \mathbb{R}^{N_{{\rm RF},r} \times N_r}$ of the analog combiner are output through a fully connected linear layer. Finally, the complex exponential function is applied to satisfy the unit modulus constraint, i.e.,
\begin{equation}
	{{\bf{W}}_{{\rm{RF}}}} = \exp \left({\rm{j}}{{\bf{\Phi }}_{{\rm{RF}}}}\right).
\end{equation}

This entire process can be summarized as
\begin{equation}
	\left\{{{\bf{W}}_{{\rm{RF}}}},{{\bf{S}}_{{\rm{CSI}},{\rm{BS}}}}\right\} =  \mathcal{C}({\bf{q}}[1], \cdots ,{\bf{q}}[K];{{\cal W}_{{\rm{BS-CSANet}}}}),
\end{equation}
where ${\mathcal W}_{\rm BS-CSANet}$ represents the learnable neural network parameters, and ${\cal C}(\cdot;{\mathcal W}_{\rm BS-CSANet})$ is the mapping function that transforms the received feedback bit vectors into the analog combiner ${{\bf{W}}_{{\rm{RF}}}}$ and the BS-side channel semantic feature $\mathbf{S}_{\rm CSI, BS}$.

{ Traditional channel features (such as angle delay/angle Doppler spectrum, correlation/covariance matrix) are high-dimensional, instantaneous, and decoupled from downstream tasks. They are sensitive to CSI acquisition overhead/noise/aging, have high acquisition and feedback costs in massive MIMO, and are not suitable for the non-orthogonal superposition and implicit DMRS designs we will introduce later. In contrast, our CSI semantic features are a low dimensional, decision aligned representation extracted by CSANet from received CSI-RS and feedback bits to directly serve the goal of semantic transmission.}

\subsection{Source and Channel Semantic Fusion at UEs}

\begin{figure*}[!t]
		\vspace*{-2mm}
	\centering
	\includegraphics[width = 2\columnwidth,keepaspectratio]{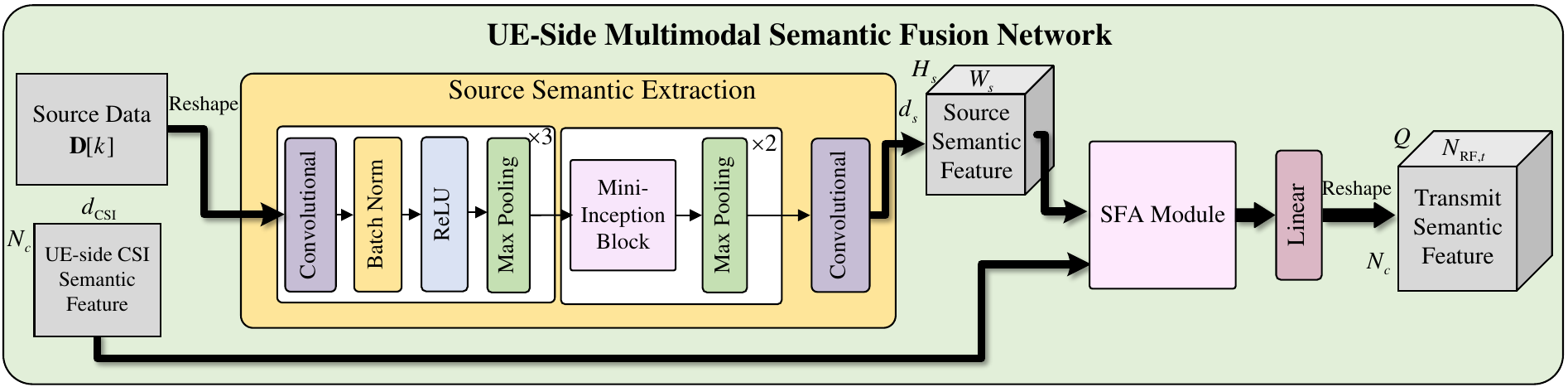}
	\captionsetup{font={footnotesize}, singlelinecheck = off, justification = raggedright,name={Fig.},labelsep=period}
	\caption{Network Architecture of the Proposed UE-MSFNet.}
	\label{fig:Fusion_UE} 
		\vspace*{-2mm}
\end{figure*}
After the design of the UE-side precoder and BS-side combiner at the physical layer, the $k$th UE needs to begin the uplink semantic transmission of the source data $\mathbf{D}[k]$. This process is modeled as the UE-MSFNet, as shown in Fig. \ref{fig:Fusion_UE}.

This paper considers the time-varying Doppler effects present in the channel, which lead to differences between the channel in the semantic transmission phase and the channel in the physical layer design phase. Additionally, the channel in each time slot of the transmission phase varies. As a result, after designing the precoder and combiner for the physical layer, the low-dimensional equivalent channel ${\mathbf{H}}_{\rm{equ}}[k,q,n] = {\mathbf{W}}_{\rm{RF}} {\mathbf{H}}[k,q,n] {\mathbf{F}}_{\rm{RF}}[k] \in \mathbb{C}^{N_{{\rm{RF}},r} \times N_{{\rm{RF}},t}}$ still changes over time. Therefore, in traditional schemes, the estimation and recovery of the low-dimensional equivalent channel at the BS receiver directly affect the final transmission performance. Specifically, the UE transmitter needs to allocate information data symbols and DMRS symbols across different resource locations. The BS receiver can estimate the equivalent channel at the resource location of the placed DMRS symbols and extrapolate the full channel information for subsequent data demodulation. However, under a fixed total transmission resource, DMRS symbols and data symbols compete for resources, creating a trade-off between them. The increase in resources required for DMRS means fewer resources are available for data transmission, resulting in a relatively low utilization of data transmission resources.

To address this issue, this paper proposes not to allocate explicit orthogonal DMRS during the semantic transmission phase. Instead, the baseband transmission signal is directly output by a neural network, which we interpret as the result of implicit DMRS allocation performed by the network. This allows for better DMRS allocation through deep learning training. Furthermore, traditional DMRS allocation schemes are typically fixed and difficult to adapt to real-time variations in channel conditions. To better adapt to different channel environments, we propose to further utilize the UE-side channel semantic feature obtained during the physical layer design phase and fuse it with the source semantic feature. This fusion helps optimize the transmission design by incorporating the underlying Doppler and other channel information.

Specifically, we first input the source data $\mathbf{D}[k]$ into a source semantic extraction network, extracting the source semantic feature with dimension $d_s\times H_s\times W_s$, where $d_s = 64$ denotes the number of feature channels, and $H_s = 10$, $W_s = 10$ represent the spatial height and width of the semantic feature, respectively. The network structure is consistent with the encoder described in \cite{sem-NOMA}, designed to extract semantics from RGB or infrared images. It primarily uses convolutional layers for feature extraction and max pooling for dimensionality reduction. 
Then, we input the source semantic features and the UE-side channel semantic feature into the proposed semantic fusion attention (SFA) module for fusion (the details of which will be presented in Subsection \ref{sec:SFA}), yielding a fused semantic feature with the same dimensions as the source semantic feature.  Finally, a fully connected layer outputs the baseband transmission semantic feature $\mathbf{S}_{\rm{BB}}[k]$ with dimensions $N_c \times N_{{\rm{RF}},t} \times Q$. The power constraint is then applied
\begin{equation}
	{{\bf{S}}_{{\rm{BB}}}}[k] = \frac{{\sqrt {{P_t}{N_c}Q} {{\bf{S}}_{{\rm{BB}}}}[k]}}{{{{\left\| {{{\bf{F}}_{{\rm{RF}}}}[k]{{\bf{S}}_{{\rm{BB}}}}[k]} \right\|}_F}}}.
\end{equation}

This entire process can be summarized as
\begin{equation}
{{\bf{S}}_{{\rm{BB}}}}[k] = {\cal E}({{\bf{S}}_{{\rm{CSI}},{\rm{UE}}}}[k], {\mathbf{D}}[k]; {\mathcal{W}}_{{\rm{UE-MSFNet}}}),
\end{equation}
where ${\mathcal{W}}_{{\rm{UE-MSFNet}}}$ represents the learnable neural network parameters, and ${\cal E}(\cdot; {\mathcal{W}}_{{\rm{UE-MSFNet}}})$ is the mapping function that transforms the source data $\mathbf{D}[k]$ and the UE-side channel semantic feature ${{\bf{S}}_{{\rm{CSI}},{\rm{UE}}}}[k]$ into the baseband transmission semantic feature $\mathbf{S}_{\rm{BB}}[k]$.

\subsection{Source and Channel Semantic Fusion at BS}

\begin{figure*}[!t]
	\centering
	\includegraphics[width = 2\columnwidth,keepaspectratio]{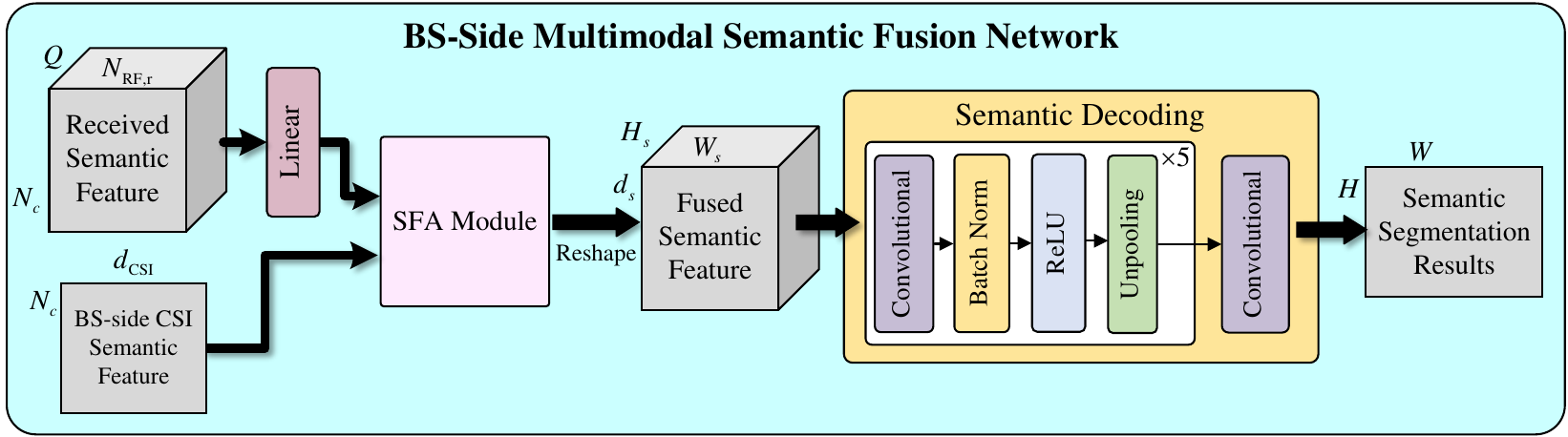}
	\captionsetup{font={footnotesize}, singlelinecheck = off, justification = raggedright,name={Fig.},labelsep=period}
	\caption{Network Architecture of the Proposed BS-MSFNet.}
	\label{fig:Fusion_BS} 
\end{figure*}

After all UEs complete the semantic transmission, the BS receives the aggregated baseband semantic signal $\mathbf{Y}_{\rm{BB}}$ with dimensions ${N_c \times N_{{\rm RF},r} \times Q}$, as shown in equations \ref{equ:rec1} and \ref{equ:rec2}. The BS needs to process this aggregated received semantic signal to obtain the final semantic task result, which is the semantic segmentation classification result $\mathbf{R} \in \mathbb{C}^{9 \times H \times W}$. We model this process as the BS-MSFNet, as depicted in Fig. \ref{fig:Fusion_BS}.

As mentioned in the previous subsection, we adopted a neural network-based implicit DMRS design at the UEs. Consequently, we apply a corresponding neural network-based semantic extraction method at the BS to avoid traditional explicit channel estimation and data demodulation. Moreover, to better adapt to different channel samples, we fuse the BS-side channel semantic feature ${{\bf{S}}_{{\rm{CSI}},{\rm{BS}}}}$ obtained during the physical layer design with the received semantic feature.

Specifically, the BS first transforms both the received semantic feature and the BS-side semantic feature through fully connected linear layers to a dimension of $d_s\times H_s\times W_s$. Then, the BS employs the SFA module to fuse the received semantic feature with the BS-side semantic feature, while preserving the dimensions as \(d_s\times H_s\times W_s\).
  Finally, the BS inputs this fused semantic feature into a semantic decoding network to produce the final semantic segmentation result. The structure of the semantic decoding network is consistent with the decoder described in \cite{sem-NOMA}, which extracts high-dimensional semantic segmentation results from low-dimensional semantic features, primarily utilizing convolutional layers for feature extraction and nearest-neighbor unpooling for upsampling.

The entire process can be expressed as
\begin{equation}
	{\bf{R}} = {\cal D}({{\bf{S}}_{{\rm{CSI,BS}}}},{{\bf{Y}}_{{\rm{BB}}}};{{\mathcal W}_{{\rm{BS-MSFNet}}}}),
\end{equation}
where ${\mathcal W}_{\rm BS-MSFNet}$ represents the learnable neural network parameters, and ${\cal D}(\cdot; {\mathcal W}_{\rm BS-MSFNet})$ is the mapping function that maps the received semantic feature ${{\bf{Y}}_{{\rm{BB}}}}$ and the BS-side channel semantic feature ${{\bf{S}}_{{\rm{CSI}},{\rm{BS}}}}$ to the baseband semantic segmentation result $\mathbf{R}$.

\begin{figure*}[!t]
		\vspace*{-2mm}
	\centering
	\includegraphics[width = 2\columnwidth,keepaspectratio]{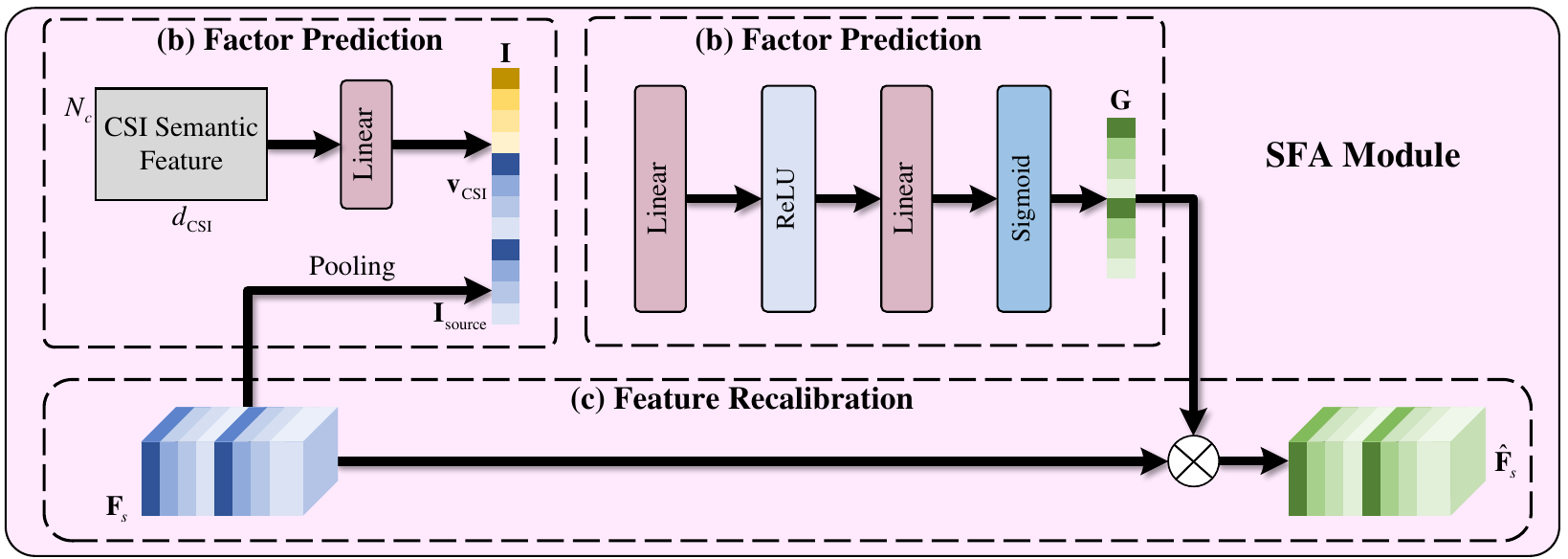}
	\captionsetup{font={footnotesize}, singlelinecheck = off, justification = raggedright,name={Fig.},labelsep=period}
	\caption{Proposed SFA module.}
	\label{fig:SFA} 
		\vspace*{-2mm}
\end{figure*}

\subsection{SFA Module}\label{sec:SFA}
We propose a SFA module for massive MIMO semantic communication systems, which jointly regulates CSI and source semantic features to enable adaptive semantic encoding and decoding under varying channel conditions. This module is inspired by the attention feature (AF) module introduced in \cite{ADJSCC}, where the original design incorporated SNR into semantic features to adapt to different SNR conditions. In our work, we extend this concept by fusing channel semantic feature with source semantic feature, thereby achieving a more refined dynamic control and feature recalibration. Overall, the proposed module considers both the influence of channel state on transmission performance and the intrinsic semantic structure of the source data, providing an efficient, robust, and adaptive solution. In the following, we describe the three main components of the proposed SFA module.

\textbf{1) Context Extraction:}  
Given the source semantic feature 	$\mathbf{F}_s \in \mathbb{R}^{d_s\times H_s\times W_s}$,
we first apply global average pooling to compute the channel-wise mean for each channel,which can be expressed as
\begin{equation}
	\mathbf{I}_{\text{source}}(i) = \frac{1}{H_s \times W_s} \sum_{j=1}^{H_s} \sum_{k=1}^{W_s} \mathbf{F}_s[i,j,k], \quad i = 1, \ldots, d_s.
\end{equation}
Concurrently, the channel semantic feature $\mathbf{F}_{\text{CSI}} \in \mathbb{R}^{N_c \times d_{\rm CSI}}$ is processed through a fully-connected layer to obtain a $d_c=16$-dimensional vector
\begin{equation}
	\mathbf{v}_{\text{CSI}} = \mathrm{FC}(\mathbf{F}_{\text{CSI}}) \in \mathbb{R}^{d_c}.
\end{equation}
The source feature vector \(\mathbf{I}_{\text{source}}\) and the CSI feature vector \(\mathbf{v}_{\text{CSI}}\) are then concatenated to form a fused context vector
\begin{equation}
	\mathbf{I} = \Big[\mathbf{I}_{\text{source}},\, \mathbf{v}_{\text{CSI}}\Big] \in \mathbb{R}^{d_s+d_c}.
\end{equation}

\textbf{2) Factor Prediction:}  
The fused context vector \(\mathbf{I}\) is fed into a factor prediction network composed of two fully-connected layers. The first layer is followed by a ReLU activation function, and the second layer is followed by a Sigmoid activation function to restrict the output to the interval \((0,1)\). This operation is formulated as
\begin{equation}
	\mathbf{G} = P_{\omega}(\mathbf{I}) = \sigma\Big(\mathbf{W}_2\, \delta\big(\mathbf{W}_1 \mathbf{I} + \mathbf{b}_1\big) + \mathbf{b}_2\Big) \in \mathbb{R}^{d_s},
\end{equation}
where \(\delta(\cdot)\) denotes the ReLU activation, \(\sigma(\cdot)\) denotes the Sigmoid activation, and \(\mathbf{W}_1\), \(\mathbf{b}_1\), \(\mathbf{W}_2\), \(\mathbf{b}_2\) are the learnable parameters. The resulting $d_s$-dimensional vector $\mathbf{G} = [G_1, G_2, \ldots, G_{d_s}]$
contains the scaling factors for each corresponding channel of the source semantic feature.

\textbf{3) Feature Recalibration:}  
Finally, the predicted scaling factors are used to recalibrate the original source semantic features on a channel-wise basis. Specifically, for the \(i\)th channel, the recalibrated feature is defined as
\begin{equation}
	\hat{\mathbf{F}}_s^i = G_i \cdot \mathbf{F}_s^i, \quad i = 1, \ldots, d_s,
\end{equation}
where \(\mathbf{F}_s^i\) denotes the \(i\)th channel of the source semantic feature. This recalibration process enables the source features to be adaptively enhanced according to the channel semantic information, thereby improving the robustness and efficiency of the semantic encoding and decoding process.

\subsection{Training Strategy}


To fully exploit the synergy between the semantic and physical layers, we adopt a three-stage training strategy that progressively pretrains individual modules and then performs joint E2E optimization {(in a sequential manner)}. {Specifically, Stage~I initializes the semantic transmission modules using the task metric (validation mIoU), Stage~II maximizes spectral efficiency to initialize BS-CSIRS-Net and UE/BS-CSANet, and Stage~III jointly fine-tunes all modules with the task loss.} This approach enhances both the semantic segmentation accuracy and the physical layer transmission efficiency. {In practice, single-stage E2E training from scratch was unstable and often failed to converge under our multi-user, massive MIMO, quantization, and non-orthogonal superposition setting; the three-stage schedule provides a stable initialization. Final model selection is based on the best validation mIoU.}

\textbf{1) Pretraining of the Semantic Transmission:}
In the first stage, the UE-MSFNet and BS-MSFNet are pretrained independently—ignoring any CSI semantic features and assuming the baseband equivalent channel is the identity matrix with no noise—to ensure a robust initialization for the semantic segmentation task. Let the final output of the semantic decoder be $	\mathbf{R} \in \mathbb{R}^{9 \times H \times W}$,
where the first dimension corresponds to 9 semantic classes. For any spatial position $(i,j)$, the predicted output is denoted by $\bigl[\mathbf{R}\bigr]_{[{:,i,j}]}$, and the corresponding one-hot label is given by $	{\mathbf y}_{\rm label}[i,j] \in \mathbb{R}^{9}$.
The training objective is defined via the cross-entropy loss
\begin{equation}
	\mathcal{L}_{\rm seg} = -\sum_{i=1}^{H}\sum_{j=1}^{W}\sum_{c=1}^{9} \left[{\mathbf y}_{\rm label}[i,j]\right]_c \log\Bigl( \bigl[\mathbf{R}\bigr]_{c,i,j} \Bigr).
\end{equation}

We evaluate segmentation performance on the validation set using the mean intersection-over-union (mIoU) metric, defined as
\begin{equation}
	\mathrm{mIoU}
	= \frac{1}{C}\sum_{c=1}^{C}
	\frac{\lvert P_c \cap G_c \rvert}
	{\lvert P_c \cup G_c \rvert},
\end{equation}
where $C=9$ is the number of classes, and $P_c$ and $G_c$ denote the sets of pixels predicted as class $c$ and the ground-truth pixels of class $c$, respectively. mIoU quantifies the average overlap between predicted and ground-truth masks across all classes. During training we save the checkpoint that achieves the highest mIoU.

\textbf{2)  Pretraining of the Physical Layer Transmission:}
In the second stage, we optimize the parameters of the BS-CSIRS-Net, BS-CSANet, and UE-CSANet modules by maximizing the spectral efficiency. Since the system involves two UEs and their semantic information is fused in subsequent stages, we neglect inter-user interference and compute the spectral efficiency for each user separately before summing them. For the $k$th UE on the $n$th subcarrier and $q$th OFDM symbols, the spectral efficiency is defined as
\begin{equation}
	\begin{split}
		\eta_{k,q,n} = \log_2 \det \Biggl( \mathbf{I} + \frac{P_t}{\sigma_n^2} \Bigl( \mathbf{W}_{\rm RF}\, \mathbf{H}[k,q,n]\, \mathbf{F}_{\rm RF}[k] \Bigr) \\
		\quad \times \Bigl( \mathbf{W}_{\rm RF}\, \mathbf{H}[k,q,n]\, \mathbf{F}_{\rm RF}[k] \Bigr)^H \Biggr).
	\end{split}
\end{equation}
where $\sigma_n^2$ is the noise power. The total spectral efficiency is then given by
\begin{equation}
	\eta = \sum_{k=1}^{2}\sum_{q=1}^{Q}\sum_{n=1}^{N_c} \eta_{k,q,n}.
\end{equation}
We define the loss function for physical layer pretraining as the negative total spectral efficiency
\begin{equation}
	\mathcal{L}_{\rm phys} = -\eta.
\end{equation}
Minimizing $\mathcal{L}_{\rm phys}$ corresponds to maximizing the spectral efficiency, thereby optimizing the design of both the precoding and combining matrices.

%

\textbf{3)  Joint E2E Training:}
After completing the above pretraining processes, the CSI-RS, UE-CSANet, BS-CSANet, UE-MSFNet, and BS-MSFNet modules are integrated for E2E joint training. {Since the ultimate objective of semantic communication is to maximize the performance of the downstream task rather than to guarantee bit-level or symbol-level recovery}, the joint training stage continues to employ the cross-entropy loss
\begin{equation}
	\mathcal{L}_{\rm joint} = \mathcal{L}_{\rm seg}.
\end{equation}
{Accordingly, the segmentation performance on the validation set is monitored via mIoU, which directly quantifies the effectiveness of the final task. The checkpoint achieving the highest mIoU is retained. This design ensures that all semantic and physical modules are coordinated to optimize the ultimate task, which aligns with the essence of semantic communication. }


{ The three-stage training strategy mirrors the roles of the two paths: the \emph{semantic transmission} path is first initialized with the task loss (Stage~1), the \emph{physical layer transmission} path is then optimized for spectral efficiency (Stage~2), and finally all modules are fine-tuned E2E on the task objective (Stage~3). This schedule ensures that channel semantics in CSANet provide a robust non-orthogonal transmission pipe, while source--channel fusion in MSFNet directly optimizes task utility under time-varying channels.}

\begin{figure*}[!t]
	\vspace{-2mm}
	\centering
	\subfigure[]{
		\begin{minipage}[t]{0.49\linewidth}
			\centering
			\label{fig:1}
			\includegraphics[width = 1\columnwidth,keepaspectratio]{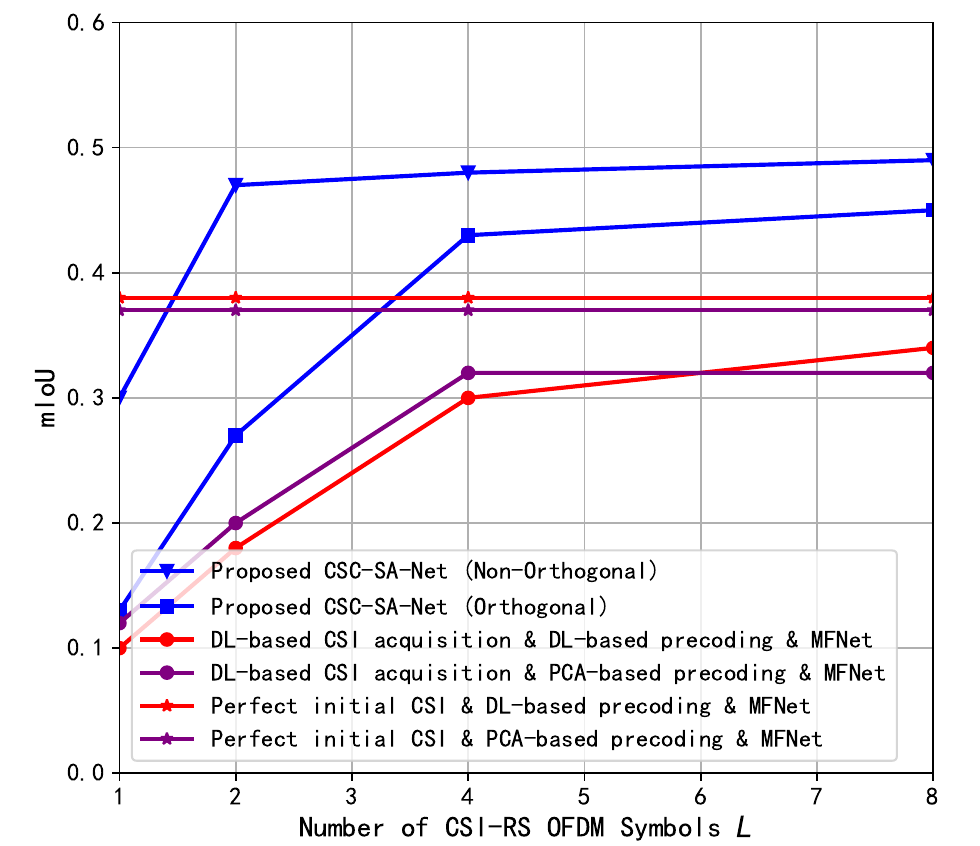}\\
		\end{minipage}%
	}%
	\subfigure[]{
		\begin{minipage}[t]{0.49\linewidth}
			\centering
			\label{fig:2}
			\includegraphics[width = 1\columnwidth,keepaspectratio]{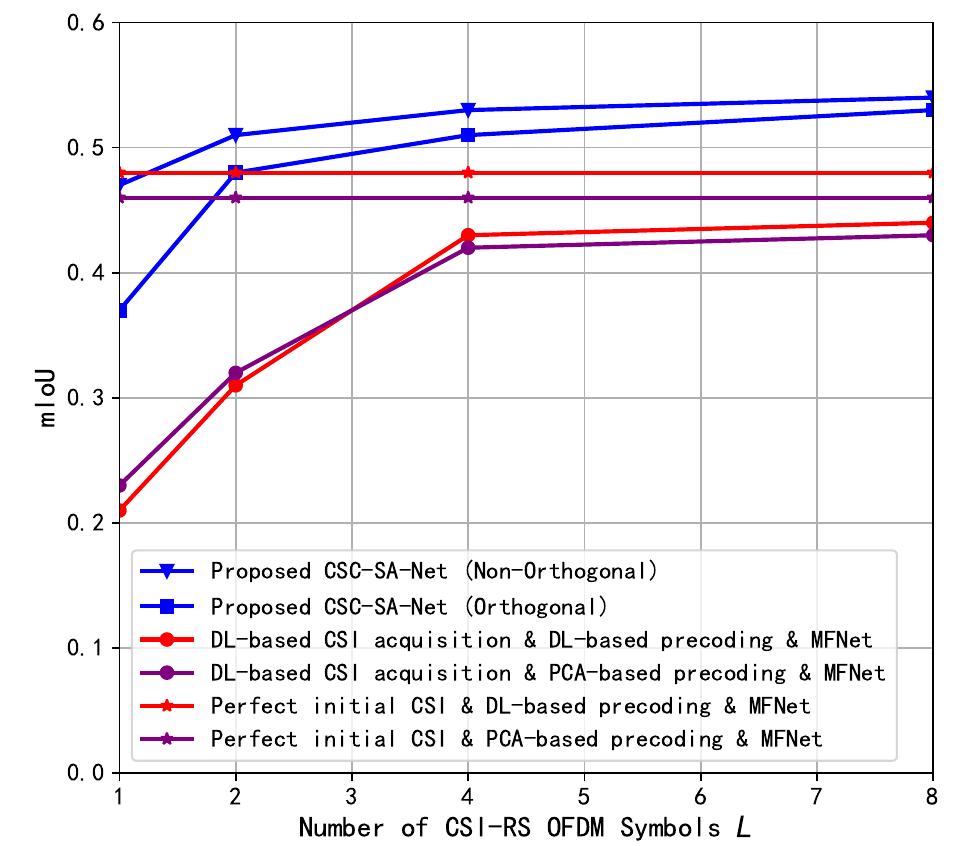}\\
		\end{minipage}%
	}%

	\centering
	\setlength{\abovecaptionskip}{-1mm}
	\captionsetup{font={footnotesize}, singlelinecheck = off, justification = raggedright,name={Fig.},labelsep=period}
	\caption{Performance comparison of different solutions versus the number of CSI-RS OFDM symbols $L$ at SNR~$=$~-25 dB, $B=1024$: (a) $Q = 1$; (b) $Q = 4$.}
	\label{fig:12} 
	\vspace*{-2mm}
\end{figure*}

\begin{figure*}[!t]
	\centering
	\subfigure[]{
		\begin{minipage}[t]{0.49\linewidth}
			\centering
			\label{fig:3}
			\includegraphics[width = 1\columnwidth,keepaspectratio]{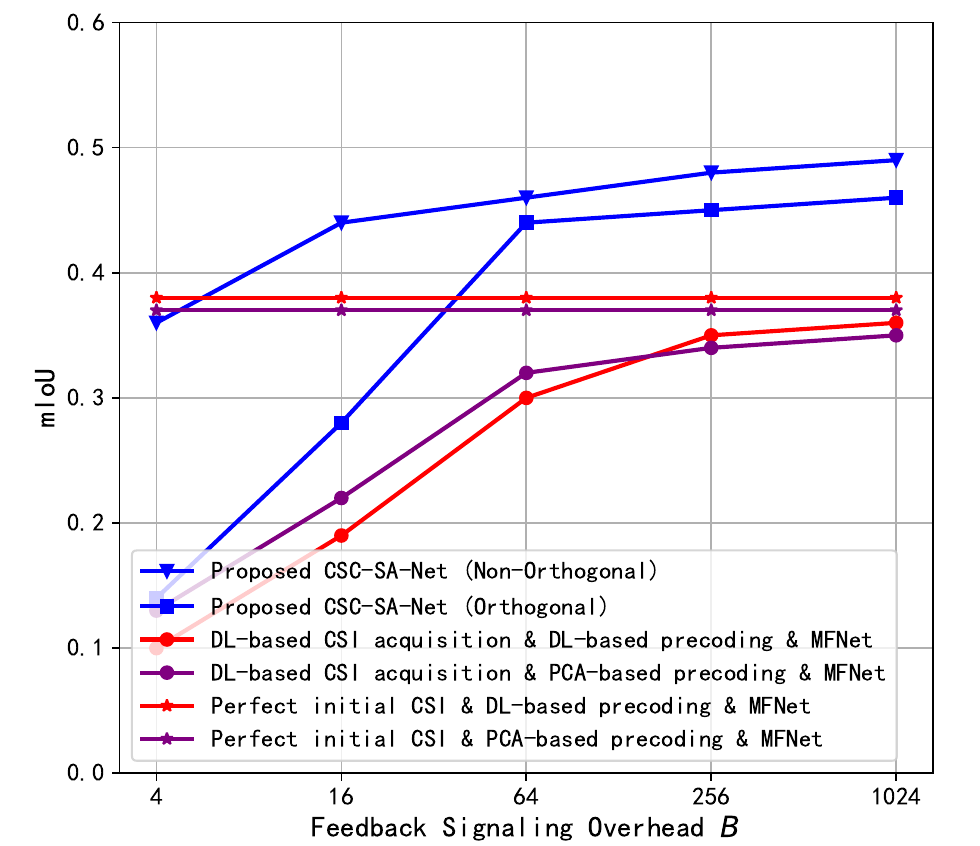}\\
		\end{minipage}%
	}%
	\subfigure[]{
		\begin{minipage}[t]{0.49\linewidth}
			\centering
			\label{fig:4}
			\includegraphics[width = 1\columnwidth,keepaspectratio]{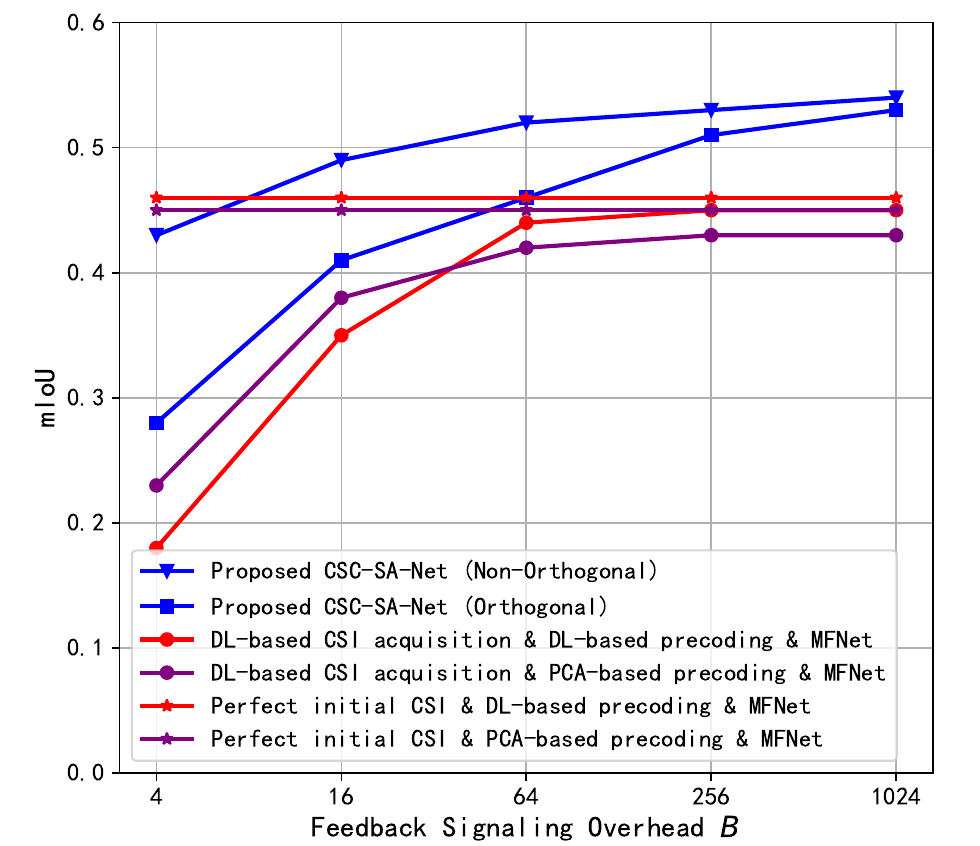}\\
		\end{minipage}%
	}%
	
	\centering
	\setlength{\abovecaptionskip}{-1mm}
	\captionsetup{font={footnotesize}, singlelinecheck = off, justification = raggedright,name={Fig.},labelsep=period}
	\caption{Performance comparison of different solutions versus the number of feedback bits $B$ at SNR~$=$~-25 dB, $L=8$ : (a) $Q = 1$; (b) $Q = 4$.}
	\label{fig:34} 
	\vspace*{-2mm}
\end{figure*}

\section{Numerical Results}\label{S:num}
\subsection{Simulation Schemes}
To evaluate the effectiveness of the proposed CSC-SA-Net framework, we compare the following schemes:
\begin{itemize}
	\item \textbf{Proposed CSC-SA-Net (Non-orthogonal):} CSC-SA-Net jointly optimizes physical-layer design and semantic transmission. In this configuration, the multi-user signals propagate through the channel and superimpose at the BS; no explicit multi-user signal detection is performed. Instead, the BS directly applies semantic fusion to execute the segmentation task.
	\item \textbf{Proposed CSC-SA-Net (Orthogonal):} We adopt the CSC-SA-Net architecture but allocate orthogonal time–frequency resources to each user to avoid inter-user interference. At the BS, the two users’ received features are concatenated and fed into the BS-side semantic fusion network for segmentation.
	\item \textbf{DL-based CSI acquisition \& DL/PCA-based precoding \& MFNet:} The BS obtains CSI via the DL-based channel estimation and feedback methods of \cite{WY_Trans}, and performs hybrid precoding using either the DL-based approach of \cite{WY_Trans} or the PCA-based method of \cite{Sun_TWC}. During semantic transmission, to mitigate Doppler-induced time variation, each user reserves one-quarter of the time–frequency grid for DMRS symbols. The BS estimates the equivalent channel from these DMRS, conducts multi-user signal demodulation, concatenates the recovered semantic features, and inputs them into the multi-spectral fusion 	networks (MFNet)  \cite{MFNet} for semantic segmentation.
	\item {\textbf{Perfect initial CSI \& DL/PCA-based precoding \& MFNet:} 
	Identical to the previous separated pipeline, except that during the CSI acquisition stage based on CSI-RS and feedback, the BS is assumed to obtain {perfect initial CSI}. Subsequent precoding and MFNet-based segmentation follow the same procedure as the baseline. Note that in the ensuing data transmission, the channel remains time-varying (i.e., Doppler effects), thus requiring DMRS-based tracking/updates; only the {initial} CSI is assumed perfect.}
	
\end{itemize}

\begin{figure*}[!t]
		\vspace{-2mm}
	\centering
	{
		\begin{minipage}[t]{0.49\linewidth}
			\centering
			
			\includegraphics[width = 1\columnwidth,keepaspectratio]{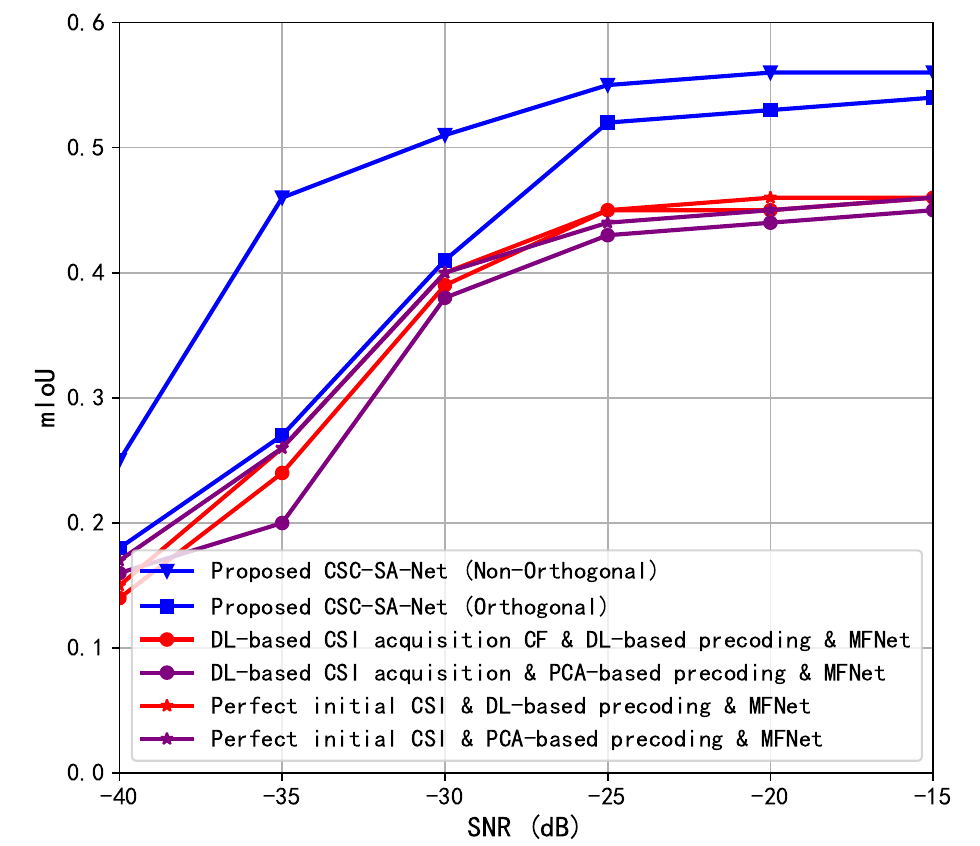}\\
			\captionsetup{font={footnotesize}, singlelinecheck = off, justification = raggedright,name={Fig.},labelsep=period}
			\caption{Performance comparison of different solutions versus SNR at $L=8$, $B=1024$, and $Q=4$.}
			\label{fig:5}
		\end{minipage}%
	}%
	{
		\begin{minipage}[t]{0.49\linewidth}
			\centering
			
			\includegraphics[width = 1\columnwidth,keepaspectratio]{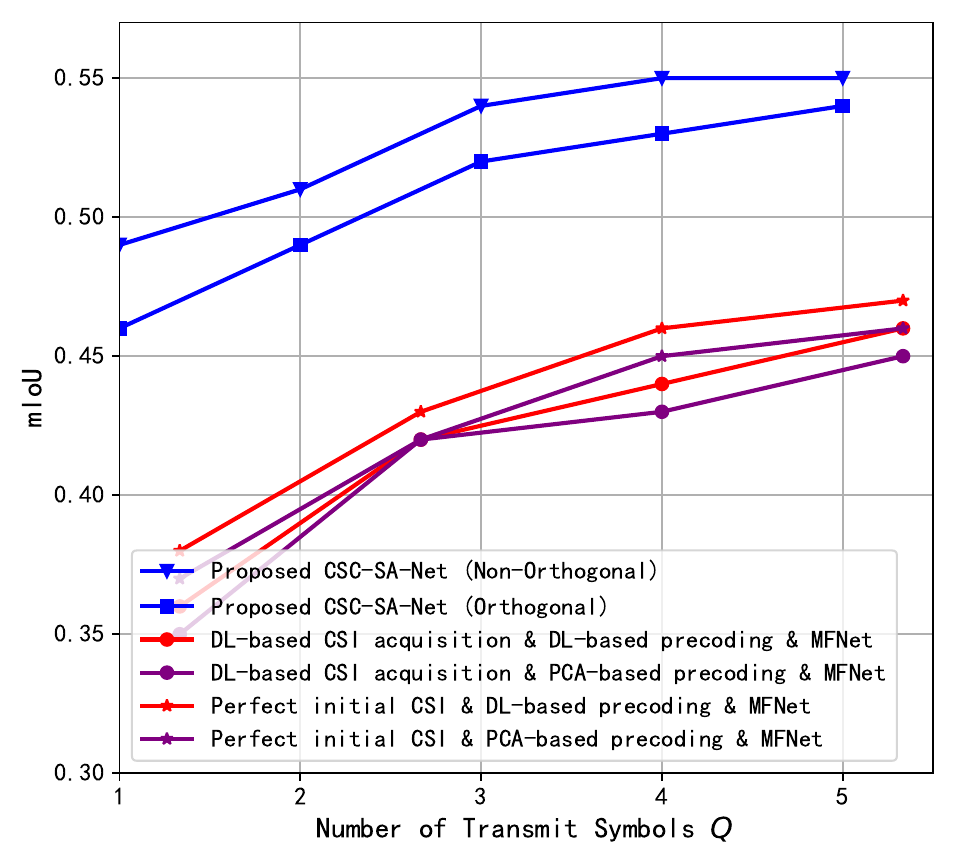}\\
			\captionsetup{font={footnotesize}, singlelinecheck = off, justification = raggedright,name={Fig.},labelsep=period}
			\caption{Performance comparison of different solutions versus the number of transmit symbols $Q$ at SNR~$=$~-25 dB, $L=8$, and $B = 1024$.}
			\label{fig:6}
		\end{minipage}%
	}%
	\vspace{-2mm}
\end{figure*}

%

\subsection{Dataset for Neural Network Training}
\paragraph{Channel Dataset}  
Channel realizations are generated according to the sparse multipath MIMO-OFDM model described in Section II. Specifically, two UEs each employ a half-wavelength ULA with $N_t=8$ antennas and $N_{{\rm RF},t}=2$ RF chains, serving $N_c=100$ subcarriers. The BS uses a ULA with $N_r=64$ antennas and $N_{{\rm RF},r}=2$ RF chains. The carrier frequency is set to 28\,GHz, and UE velocities are uniformly distributed in the range 0--120\,km/h.

\paragraph{Semantic Segmentation Dataset}  
For the segmentation task, we use the RGB-Thermal Dataset introduced in \cite{MFNet}, which comprises 1569 aligned RGB–thermal image pairs of resolution $480\times640$, covering diverse daytime and nighttime urban road scenes with pixel-level annotations for eight classes (car, person, bicycle, road curvature, parking sign, guardrail, traffic cone, speed bump). The dataset is split temporally into a training set (50\% daytime + 50\% nighttime), a validation set (25\% + 25\%), and a test set (25\% + 25\%).

\subsection{Training Settings}
All models are implemented in PyTorch and trained on a workstation equipped with two NVIDIA GeForce RTX 4090D GPUs. {{For the training of the proposed CSC-SA-Net (both non-orthogonal and orthogonal modes), we employ the Adam optimizer and follow the three-stage training strategy described in Section~III-H. The batch size is set to 16 in the first and third stages, and 128 in the second stage. The learning rate is fixed at $10^{-4}$ across all stages. For the baseline schemes, both the DL-based CSI acquisition and DL-based precoding modules are trained with a batch size of 128 and a learning rate of $10^{-4}$. For the MFNet baseline, we use a batch size of 16 and the same learning rate of $10^{-4}$. In all schemes, we apply data augmentation by random image flipping to enlarge the dataset and mitigate overfitting, and adopt an early stopping strategy to save the best-performing checkpoint.}}

\subsection{Performance Comparison}

Fig.~\ref{fig:12} illustrates the mIoU performance of the semantic segmentation task versus the number of CSI-RS OFDM symbols for each scheme. When channel estimation is inaccurate ($L<4$), both variants of the separated design—``DL-based CSI acquisition \& DL/PCA-based precoding \& MFNet''—exhibit severe performance degradation due to the large mismatch between actual and estimated CSI. This mismatch leads to poor received signal quality at the BS and consequently a significant drop in the segmentation accuracy of the MFNet decoder. Even with a sufficient CSI-RS ($L\ge4$), these separated schemes remain bottlenecked: under perfect initial CSI assumptions (``Perfect initial CSI \& DL/PCA-based precoding \& MFNet''), their mIoU still falls well below that of the proposed CSC-SA-Net. This is because the effective channel after independently optimized precoders and combiners deviates from a simple AWGN model, and an MFNet decoder trained in isolation cannot adapt to such complex channel characteristics {(i.e., the link-level optimization goal is not aligned with the task-level objective)}. In contrast, the proposed CSC-SA-Net’s E2E training allows the network to learn and compensate for the actual channel effects {by aligning the physical-layer design with the downstream task (mIoU) and jointly optimizing source–channel semantics.}

{Furthermore, the advantage of CSC-SA-Net is amplified in the CSI-RS-limited regime ($L<4$): the separated schemes must spend additional resources to maintain CSI fidelity for each user and stage, whereas task-aligned E2E learning exploits channel semantics to remain robust under CSI under-specification and reduces the propagation of estimation errors across modules. Even when $L\!\ge\!4$, the separated design still suffers from suboptimal task utility because its independently trained decoder cannot absorb the residual, non-AWGN channel distortions introduced by the precoder/combiner pair.}

Moreover, even with perfect initial CSI in the channel estimation and feedback stages, the semantic transmission phase retains strong Doppler-induced time variations, making the low-dimensional equivalent channel time-varying and necessitating explicit DMRS insertion. {Under such fast-fading conditions, the overhead associated with explicit DMRS limit the amount of task-useful symbols, which further widens the gap between link-level optimality and task-level performance in the separated schemes.} These results highlight that traditional explicit DMRS allocation is insufficient for fast-varying channels. The proposed CSC-SA-Net avoids explicit DMRS insertion during semantic transmission by using an implicit, data-driven DMRS allocation mechanism, thereby eliminating data–DMRS resource competition and adapting to time-varying channels for enhanced performance {while preserving more symbols for task-relevant semantic features}.

Fig.~\ref{fig:1} compares the two CSC-SA-Net variants under highly constrained transmission resources ($Q=1$). The proposed CSC-SA-Net (non-orthogonal) achieves a clear mIoU advantage over the proposed CSC-SA-Net (orthogonal), particularly at low CSI-RS budgets ($L<4$). {This gain is amplified when both the effective data symbol budget and CSI-RS resources are scarce, where superposition yields a higher task-useful information rate.} This advantage stems from the proposed CSC-SA-Net’s ability to non-orthogonally superimpose multi-user semantics and perform direct fusion, whereas the orthogonal scheme must divide time–frequency resources equally between users, compressing each user’s semantic features more severely and limiting performance. {The non-orthogonal scheme better exploits cross-user correlation, while the orthogonal scheme's per-user orthogonalization incurs higher overhead and weaker cross-user information aggregation.} Since all users share the same semantic task, channel superposition naturally implements semantic fusion without incurring additional orthogonal transmission overhead. In Fig.~\ref{fig:2}, where transmission resources are more ample ($Q=4$), the orthogonal scheme can allocate sufficient capacity to each user’s semantics, reducing the performance gap between the two approaches.

Fig.~\ref{fig:34} presents the mIoU performance versus feedback signaling overhead. Consistent with the CSI-RS budget results, at low feedback bit budgets ($B<64$), the separated-design schemes suffer catastrophic mIoU loss due to CSI estimation errors. Meanwhile, the CSC-SA-Net’s non-orthogonal multi-user fusion strategy again outperforms its orthogonal counterpart, with the most pronounced gains observed under low transmission symbol budgets ($Q=1$).

\begin{figure*}[!t]
		\vspace*{-1mm}
	\centering
	\includegraphics[width = 2 \columnwidth,keepaspectratio]
	{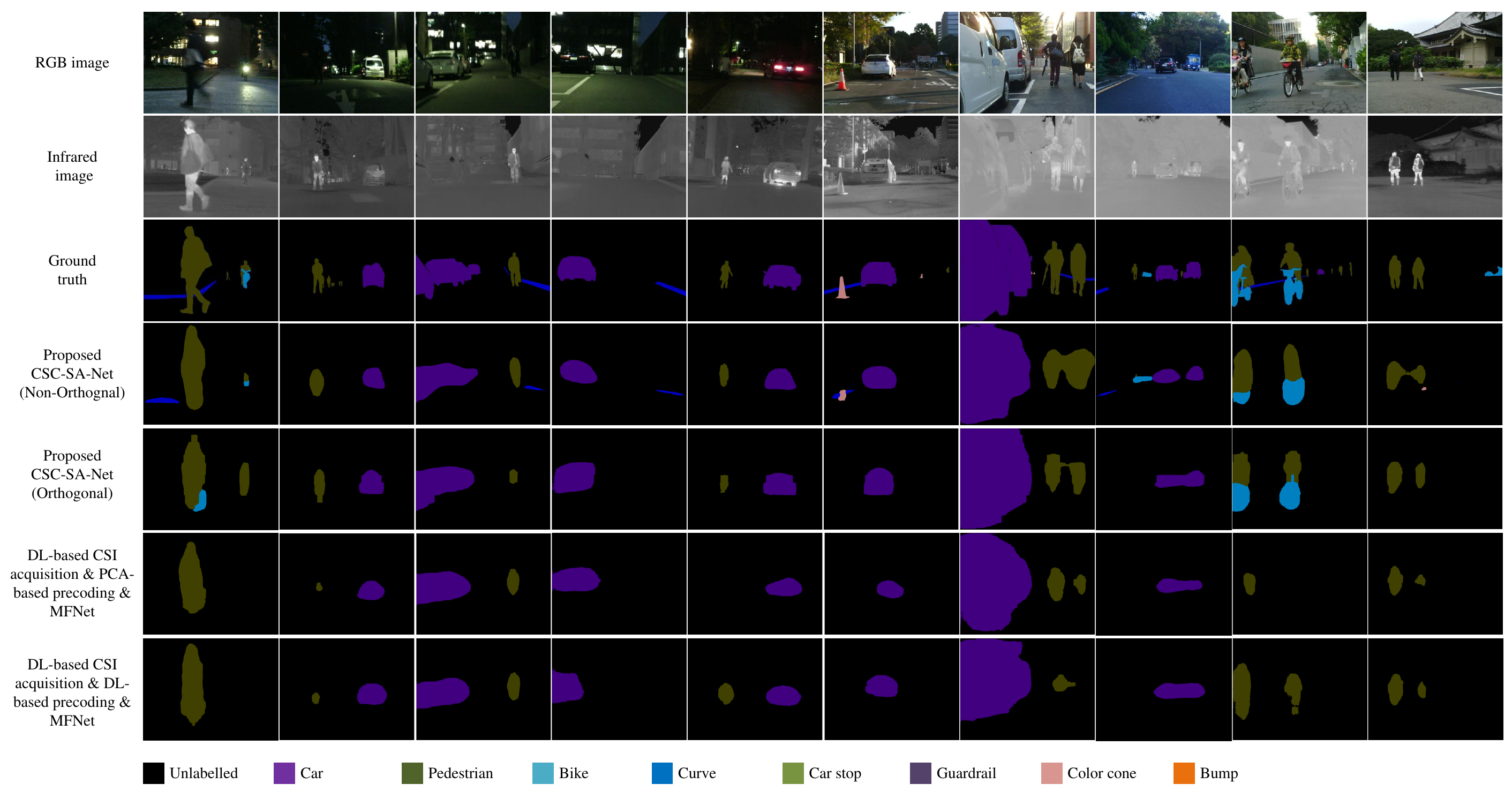}
	\captionsetup{font={footnotesize}, singlelinecheck = off, justification = raggedright,name={Fig.},labelsep=period}
	\caption{Examples of prediction for different schemes at SNR~$=$~-25 dB, $L=8$, $Q=4$, and $B = 1024$.}
	\label{fig:visual} 
	\vspace*{-1mm}
\end{figure*}

Fig.~\ref{fig:5} shows the mIoU performance of each scheme versus SNR. At high SNR, both the CSC-SA-Net variants (orthogonal and non‑orthogonal) markedly outperform the separated designed baselines. This improvement arises because the CSC-SA-Net’s E2E training jointly optimizes implicit DMRS placement and CSI‑semantic feature extraction, allowing the network to learn and compensate for residual Doppler‑induced CSI mismatches.

Under low‑SNR conditions, the non‑orthogonal CSC-SA-Net yields additional gains over its orthogonal counterpart. By superimposing multiple users’ semantic embeddings in the analog domain, the CSC-SA-Net’s non‑orthogonal transmission realizes an implicit combining of correlated semantic features, thereby increasing the effective SNR of the aggregated semantic representation.  In contrast, orthogonal transmission treats each user's semantics independently, preventing any constructive combination and rendering each stream more susceptible to channel noise. As a result, the non‑orthogonal CSC-SA-Net achieves significantly higher segmentation accuracy at low SNR.

Fig.~\ref{fig:6} compares the mIoU performance of the semantic segmentation task versus the number of transmit symbols for each scheme{\footnote{ For baselines that reserve {explicit} DMRS (25\% overhead), we equalize semantic payload by mapping \(Q_{\mathrm{base}}=\tfrac{4}{3}Q\). When \(Q\) is not a multiple of 3, \(Q_{\mathrm{base}}\) is implemented via partial-symbol loading (fractional subcarriers in the last symbol), which causes the baseline curves to appear off the integer ticks.}}. All schemes exhibit steadily increasing segmentation accuracy with more transmit symbols, owing both to the ability to convey additional semantic content and to the increased redundancy that improves resilience to channel noise. Moreover, the proposed CSC-SA-Net—under both orthogonal and non‑orthogonal transmission—consistently and significantly outperforms the separated designed baselines. This superior performance arises from the CSC-SA-Net’s E2E training, which enables the network to adapt more effectively to non‑ideal, time‑varying channel conditions.

Fig.~\ref{fig:visual} presents representative semantic segmentation outputs for each scheme. The proposed CSC-SA-Net, under both orthogonal and non‑orthogonal transmission, accurately delineates multiple semantic classes, although object boundaries can appear slightly blurred due to implicit channel distortions and noise in the E2E inference process. By comparison, the baseline separated design yields noticeably coarser segmentation masks.

{{Note that this paper focuses on non-orthogonal semantic fusion for multiple users collaboratively serving the same task. Our experiments instantiate this framework on a semantic segmentation task using an RGB--Thermal dataset. To target other modalities and tasks, it suffices to replace the {source semantic extraction} module in the UE-side multimodal semantic fusion network according to the modality, and to replace the {semantic decoding} head in the BS-side multimodal semantic fusion network according to the task. In this way, the framework readily transfers to image stitching, VQA, object detection/tracking, multimodal perception for autonomous driving, and related cooperative tasks. Detailed designs and training paradigms for additional modalities and tasks are left for future work.}}

\begin{table}[t]
	\centering
	\footnotesize
	\setlength{\tabcolsep}{4pt}
	\captionsetup{font={color = {black}}}
	\caption{Complexity and Memory of Compared Schemes}
	\label{tab:comp_mem}
	
	\begin{tabularx}{\columnwidth}{@{}>{\raggedright\arraybackslash}X
			>{\raggedleft\arraybackslash}c
			>{\raggedleft\arraybackslash}c
			>{\raggedleft\arraybackslash}c@{}}
		\toprule
		\textbf{Schemes} & \textbf{FLOPs} & \textbf{Parameters} & \textbf{Peak GPU Mem } \\
		\midrule
		Proposed CSC-SA-Net (Non-Orthogonal; physical-layer CSI semantic extraction modules) & 155M & 45.8M & 4.2M \\
		Proposed CSC-SA-Net (Non-Orthogonal; multimodal semantic fusion modules)             & 12G  & 3.37M & 58.67M \\
		Proposed CSC-SA-Net (Non-Orthogonal)                                                 & 12G  & 50.1M & 61.52M \\
		Proposed CSC-SA-Net (Orthogonal)                                                     & 12G  & 50.1M & 61.4M  \\
		DL-based CSI acquisition                                                              & 425M & 849M  & 5.78M  \\
		DL-based precoding                                                                    & 661M & 35.8M & 6.24M  \\
		MFNet                                                                                 & 12G  & 3.23M & 58.63M \\
		\bottomrule
	\end{tabularx}
\end{table}

{ \subsection{Complexity Analysis}
	To evaluate the practicality of the proposed framework, we analyze the computational and memory costs of both the baseline schemes and CSC-SA-Net. We report three quantifiable metrics: (i) model parameter counts, (ii) FLOPs per inference, and (iii) peak GPU memory consumption. The results are summarized in Table~\ref{tab:comp_mem}.
	
	For the separated baseline designs (DL-based CSI acquisition, DL-based precoding, and MFNet), the majority of the computational burden and GPU memory footprint is attributed to MFNet, which extracts and decodes high-dimensional RGB–thermal source features for semantic segmentation. MFNet requires up to 12G FLOPs per inference and 58.6M GPU memory per sample, whereas the physical-layer modules handling CSI features incur much smaller costs due to the low dimensionality of CSI. The parameter count of MFNet remains modest (3.23M) because of its CNN-based architecture.
	
	For the proposed CSC-SA-Net, we separately report the costs of (i) the physical-layer CSI semantic extraction modules (BS-CSIRS-Net, UE-CSANet, BS-CSANet) and (ii) the multimodal semantic fusion modules (UE-MSFNet and BS-MSFNet). Similar to the baselines, the main overhead originates from the fusion part, which processes RGB–thermal data. The CSI-related modules consume only a small fraction of the resources. Moreover, the complexity remains nearly identical under both NOMA and OMA configurations.
	
	Overall, the proposed CSC-SA-Net introduces no significant extra computational or memory overhead compared with separated designs. Both approaches are dominated by the cost of source feature processing, yet CSC-SA-Net achieves substantial performance improvements in semantic segmentation accuracy and robustness, confirming its effectiveness and practicality.}

\section{Conclusion}


In this paper, we proposed the CSC-SA-Net, an E2E framework for multimodal semantic non-orthogonal transmission and fusion in hybrid massive MIMO-OFDM systems. By jointly optimizing CSI-RS design, analog beamforming, and semantic transmission, the proposed CSC-SA-Net unifies physical-layer and semantic-level objectives. The proposed CSC-SA-Net eliminates explicit DMRS allocation via E2E learning and leverages non-orthogonal multi-user transmission to fuse semantic features directly in the analog domain, enhancing transmission efficiency. The SFA mechanism further enhances semantic fusion by adaptively merging CSI and source semantic features. Extensive experiments on RGB–thermal segmentation demonstrate that CSC-SA-Net achieves superior performance over conventional baselines, particularly in resource-constrained scenarios.

\bibliographystyle{IEEEtran}

\end{document}